\documentclass[10pt,journal,compsoc]{IEEEtran}
\ifCLASSOPTIONcompsoc
  \usepackage[nocompress]{cite}
\else
  \usepackage{cite}
\fi
\usepackage{hyperref}
\hypersetup{
    colorlinks=true,
    linkcolor=blue,
    citecolor=blue
}
\usepackage{cancel}
\usepackage[dvipsnames]{xcolor}
\usepackage{dblfloatfix}
\usepackage[pdftex]{graphicx}
\usepackage{booktabs}
\usepackage{amssymb}
\usepackage{makecell}
\usepackage{multirow}
\usepackage{xcolor}
\usepackage{pifont}
\usepackage{amsmath}
\usepackage{multicol}
\usepackage{wrapfig}
\usepackage{subcaption}
\usepackage{lstautogobble}
\usepackage{dblfloatfix}
\usepackage{listings}
\usepackage{url}
\usepackage[normalem]{ulem}
\usepackage{flushend}
\usepackage[linesnumbered,ruled,vlined]{algorithm2e}

\SetCommentSty{mycommfont}
\def\tool{\textsc{Holmes}\xspace}
\definecolor{dkgreen}{rgb}{0,0.6,0}
\definecolor{gray}{rgb}{0.5,0.5,0.5}
\definecolor{mauve}{rgb}{0.58,0,0.82}
\definecolor{dblue}{rgb}{0,0,0.8}
\newcolumntype{P}[1]{>{\centering\arraybackslash}p{#1}}
\newcolumntype{M}[1]{>{\centering\arraybackslash}m{#1}}
\usepackage{footnote}
\makesavenoteenv{tabular}
\makesavenoteenv{table}
\begin{document}
\title{Modeling Functional Similarity in Source Code with Graph-Based Siamese Networks}
\author{Nikita~Mehrotra,
        Navdha~Agarwal,
        Piyush~Gupta,
        Saket~Anand,
        David~Lo,
        and~Rahul~Purandare
\IEEEcompsocitemizethanks{\IEEEcompsocthanksitem N. Mehrotra, N. Agarwal,
P. Gupta, 
S. Anand, and 
R. Purandare are with the Department
of Computer Science Engineering, IIIT Delhi, India (e-mail: nikitam@iiitd.ac.in, navdha16250@iiitd.ac.in, piyush16066@iiitd.ac.in, anands@iiitd.ac.in, purandare@iiitd.ac.in).
\IEEEcompsocthanksitem D. Lo is with the School of Information Systems
SMU, Singapore (e-mail: davidlo@smu.edu.sg)}
\thanks{}}
\IEEEtitleabstractindextext{%
\begin{abstract}
Code clones are duplicate code fragments that share (nearly) similar syntax or semantics. Code clone detection plays an important role in software maintenance, code refactoring, and reuse. A substantial amount of research has been conducted in the past to detect clones. A majority of these approaches use lexical and syntactic information to detect clones. However, only a few of them target semantic clones. Recently, motivated by the success of deep learning models in other fields, including natural language processing and computer vision, researchers have attempted to adopt deep learning techniques to detect code clones. These approaches use lexical information (tokens) and(or) syntactic structures like abstract syntax trees (ASTs) to detect code clones. However, they do not make sufficient use of the available structural and semantic information hence, limiting their capabilities.
\par This paper addresses the problem of semantic code clone detection using program dependency graphs and geometric neural networks, leveraging the structured syntactic and semantic information. We have developed a prototype tool \tool, based on our novel approach and empirically evaluated it on popular code clone benchmarks. Our results show that \tool performs considerably better than the other state-of-the-art tool, TBCCD. We also evaluated \tool on unseen projects and performed cross dataset experiments to assess the generalizability of \tool.  
Our results affirm that \tool outperforms TBCCD since most of the pairs that \tool detected were either undetected or suboptimally reported by TBCCD.
\end{abstract}
}
\maketitle

\IEEEdisplaynontitleabstractindextext
\IEEEpeerreviewmaketitle
\IEEEraisesectionheading{\section{Introduction}\label{sec:introduction}}
\IEEEPARstart{C}{ode} clones are code fragments that are similar according to some definition of similarity \cite{738528}. There are two types of similarity defined between code snippets: 1) syntactic (textual) similarity and 2) semantic similarity. Syntactic clones are code pairs that have similar syntactic structure. They share similar (or nearly similar) program text, control flow, data flow, and data-types. Semantic clones are syntactically dissimilar code snippets that share similar functionality \cite{Roy07asurvey}. Figure \ref{fig:semantic code pair} shows an example of semantic clones that sorts an array of natural numbers.
\begin{figure}[t]
\begin{lstlisting}[caption=$Sort_{1}.java$,label=code1,captionpos=b,firstnumber=1]
public static void main(String[] args){
    Scanner in = new Scanner(System.in) ;
    int T = in.nextInt() ; 
    int[] a = new int[T] ;
    for (int j = 0 ; j < T ; j ++) 
        a[j] = in.nextInt() ;
    int c = 0 ;
    for (int j = 0 ; j < T ; j ++)
        if (a[j] == j+1)
            c ++ ;
    System.out.println("Case #"+i+":"+((double)T-(double)c));
}

\end{lstlisting}
\begin{lstlisting}[captionpos=b,caption=$Sort_{2}.java$,label=code2,firstnumber=1]
public static void main(String[] args) {
    Scanner in = new Scanner(System.in);
    int n = in.nextInt(),t=0;
    float count=0.0f;
    while(n>0){
        if(++t!=in.nextInt())
            count++;
        n--;
    }
    Formatter formatter = new Formatter();
    System.out.println(formatter.format("Case#"+i+":
    %.6f",count));
}
 \end{lstlisting}
\caption{A semantic code clone example detected by HOLMES, which was reported as false negative by TBCCD. The code in Listings \ref{code1} and \ref{code2} sort an array of natural numbers by randomly shuffling the array n times.}
    \label{fig:semantic code pair}
\end{figure}
\begin{figure}[t]
    \centering
    \includegraphics[width=7cm,height=6cm]{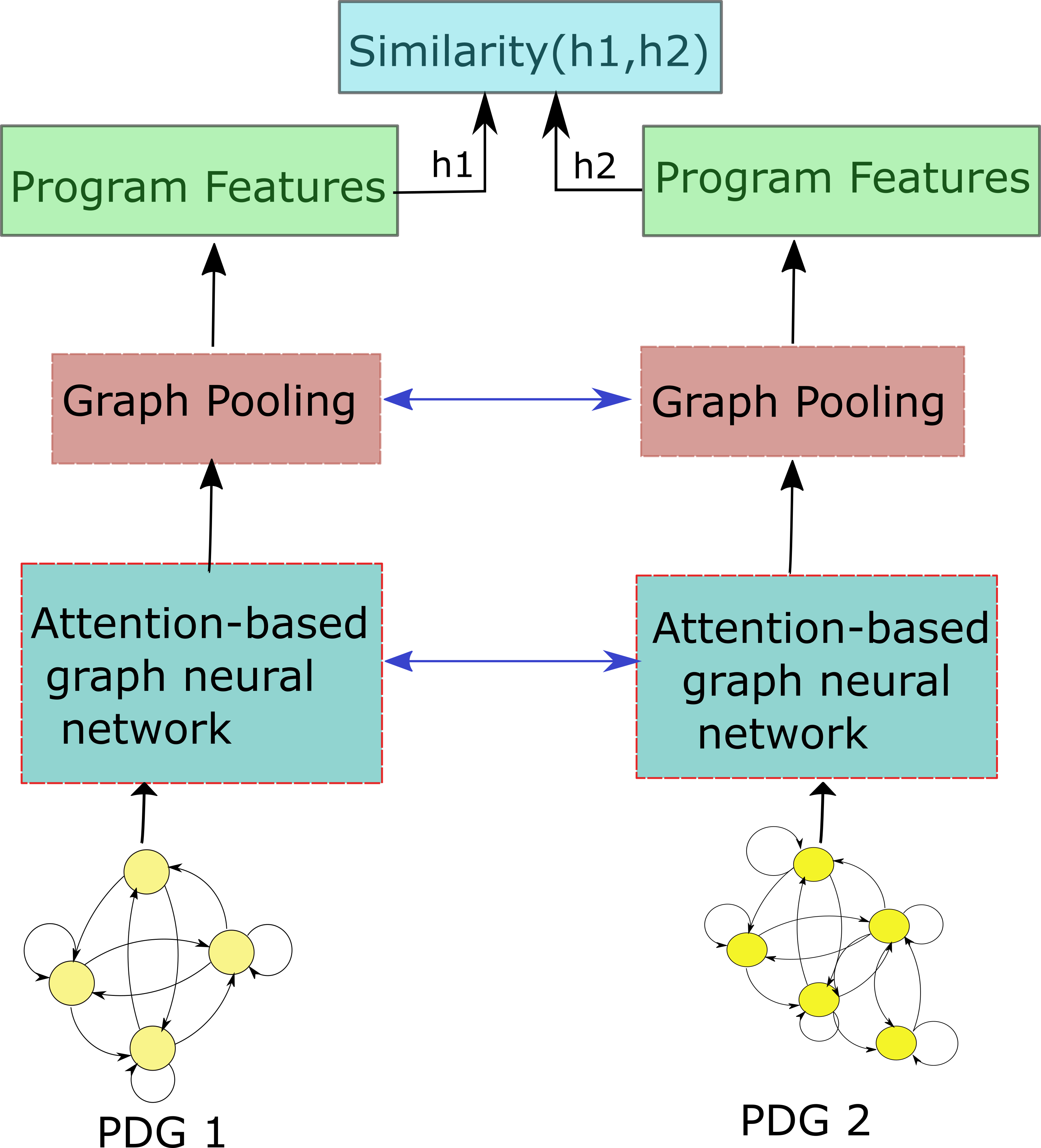}
    \caption{The proposed Siamese deep neural network consists of two identical sub-networks.  The input to the subnetwork is the pair of Java methods represented as PDGs. Each subnetwork incorporates an attention-based graph neural network model to learn PDGs node features, which are then aggregated using soft attention to constitute a graph-level representation of the given input method. The proposed network is trained to learn the similarity between two feature vectors h1 and h2. Horizontal blue arrows denote that the two sub-networks share the same set of weights and parameters.}
    \label{fig:approach overview}
\end{figure}
\par Code clones materialize in a software project when developers reuse the existing code by \textit{copy-paste-modify} operation or when they re-implement an already existing similar functionality \cite{Roy07asurvey, 10.1145/2970276.2970326}. Code clones can lead to increased software maintenance costs \cite{Thummalapenta, 7880507}. They may complicate software evolution as bug fixes and changes have to be propagated to all the clone locations \cite{Li2004CPMinerAT,4976382,1011328,738528}. However, clones are not always disastrous \cite{4023973}. They can aid in code search \cite{10.1145/2568225.2568292}, refactoring \cite{10.1145/1985404.1985423}, and bug detection \cite{10.1145/1287624.1287634}. 
\par A substantial amount of research effort has been put in to detect and analyze syntactic clones. These techniques \cite{738528,6032469, 4222572} use various handcrafted lexical and syntactic program features to identify similar (clone) pairs. However, in recent years with the growing research efforts into applying deep learning techniques for software engineering problems, researchers have adopted deep learning 
models to detect software clones \cite{10.1109/ICPC.2019.00021,10.1145/3236024.3236068,10.5555/3172077.3172312,10.1145/2970276.2970326,ijcai2018-394,Wang2020DetectingCC,7582748,8595238,8094426}. The code clone detection process begins by modeling the functional behaviors of the source code. To achieve this goal, the program features defining the source code's functionality are learned. Diverse program representations comprising of tokens, Abstract Syntax Trees (ASTs), Control Flow Graph (CFGs), Data Flow Graphs (DFGs) are being used to learn program features. For instance, Yu et al. \cite{10.1109/ICPC.2019.00021} used tree-based convolutions that exploit structural and lexical information from the ASTs of the code fragments. Notwithstanding this, we argue that these program representations do not capture program semantics even though it might be crucial for measuring code functional similarity. Thus, a more sophisticated program representation is required to learn the functional behaviors of source code.
\par A few techniques exploit Program Dependence Graphs (PDGs) for measuring code functional similarity. These techniques construct program dependence graphs for each code snippet and use graph isomorphism to measure code functional similarity. For instance, Krinke \cite{957835} used PDG representation of code snippets and modeled clone detection problem as \textit{maximal similar subgraph construction} problem. Gabel et al. \cite{10.1145/1368088.1368132} compared program dependence graphs of code pairs to detect clones. They reduced the graph isomorphism problem to a tree matching problem by mapping PDG representation to AST. However, the techniques are imprecise and are not scalable in practice due to the inherent complexity of graph isomorphism and the approximations made while mapping PDG's subgraphs to ASTs in \cite{10.1145/1368088.1368132}. 

\begin{figure*}[!b]
    \begin{minipage} {.5\textwidth}
    \centering
          \includegraphics[width=6.5cm,height=5.6cm]{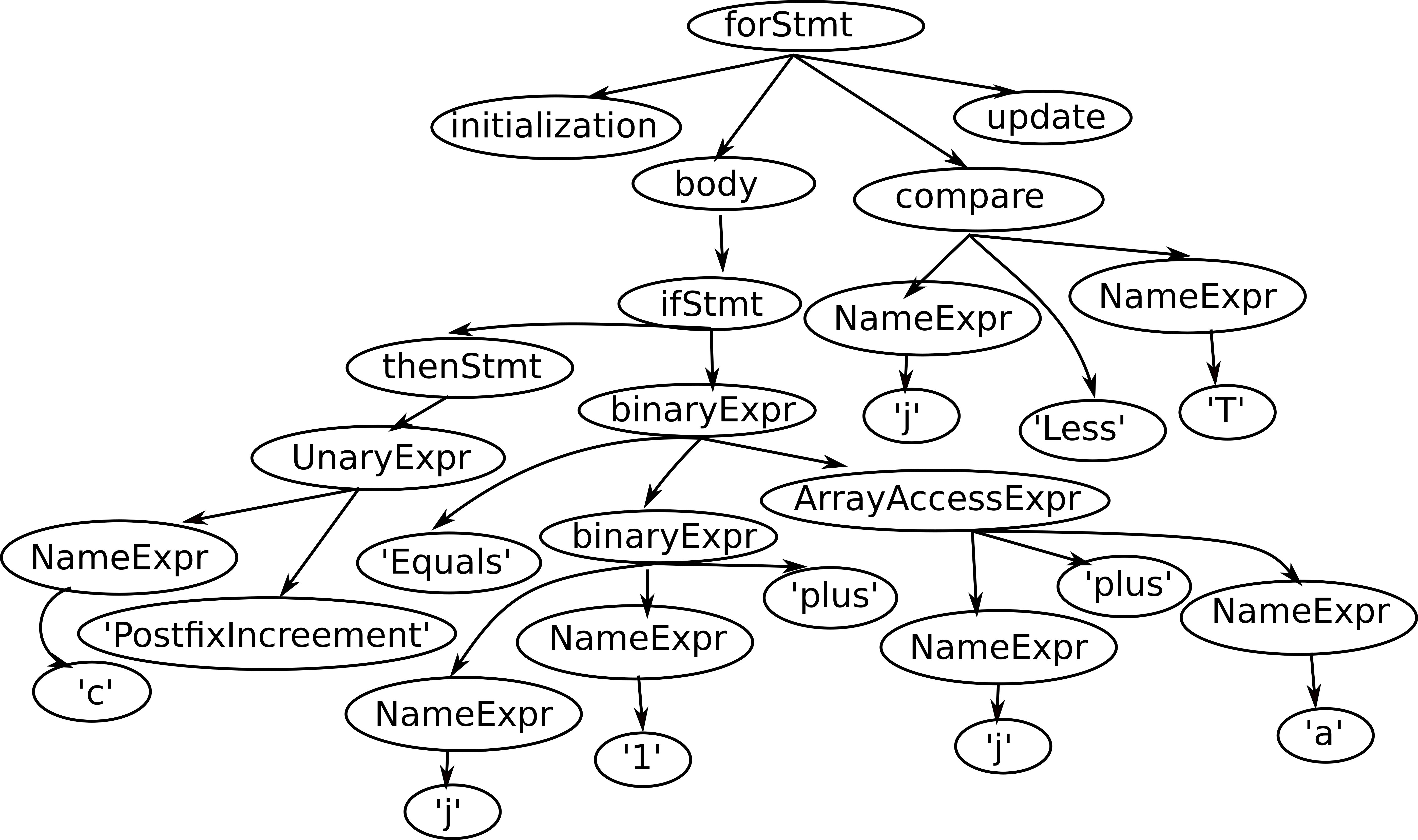}
              \caption{AST for Listing \ref{code1}.}
        \label{AST1}
    \end{minipage}
      \begin{minipage} {.5\textwidth}
    \centering
          \includegraphics[width=6.5cm,height=5.6cm]{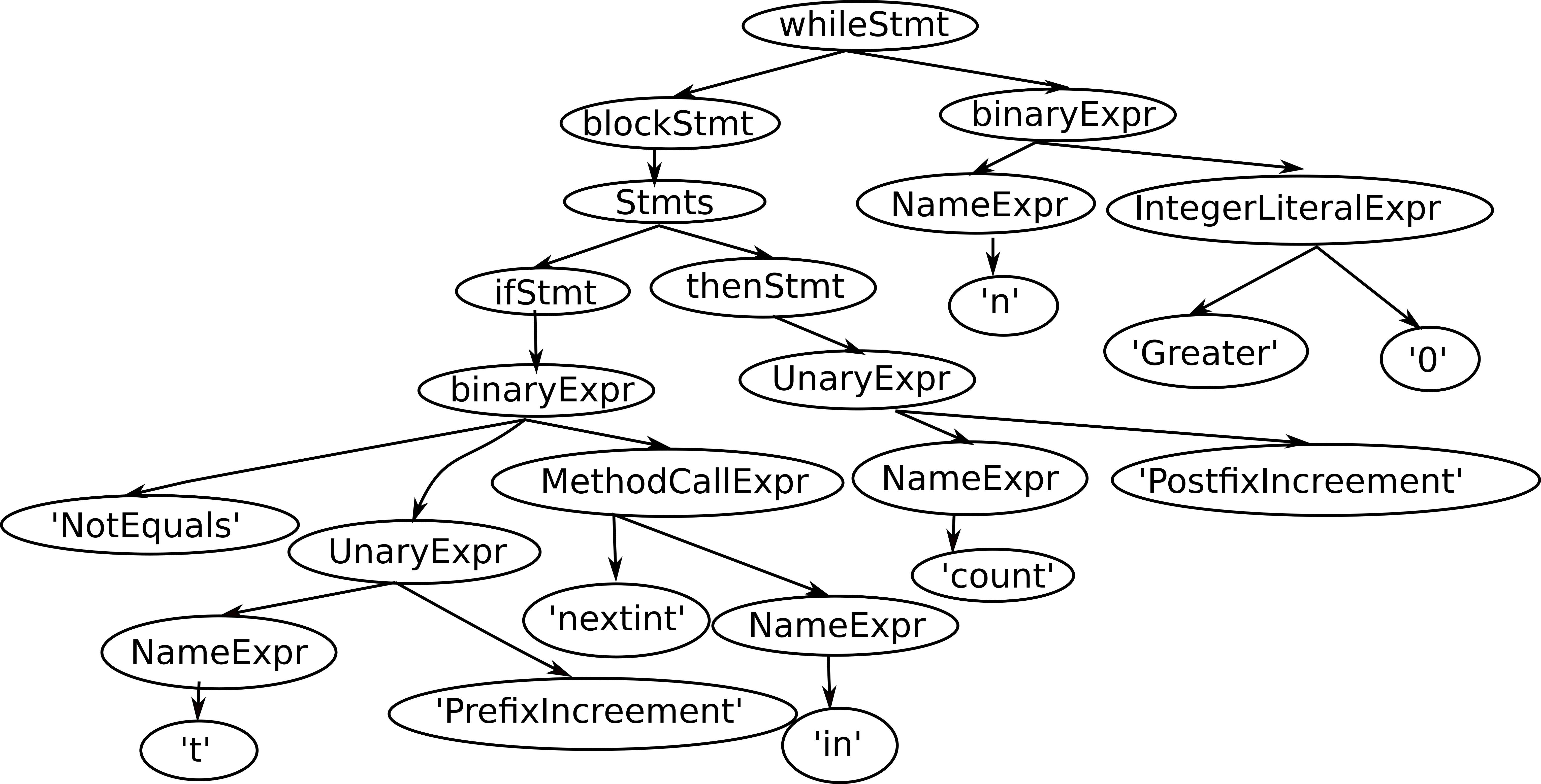}
              \caption{AST for Listing \ref{code2}.}
        \label{AST2}
    \end{minipage}
\end{figure*}
\begin{figure*}[!b]
    \begin{minipage}{.5\textwidth}
    \centering
          \includegraphics[width=6.5cm,height=5.5cm]{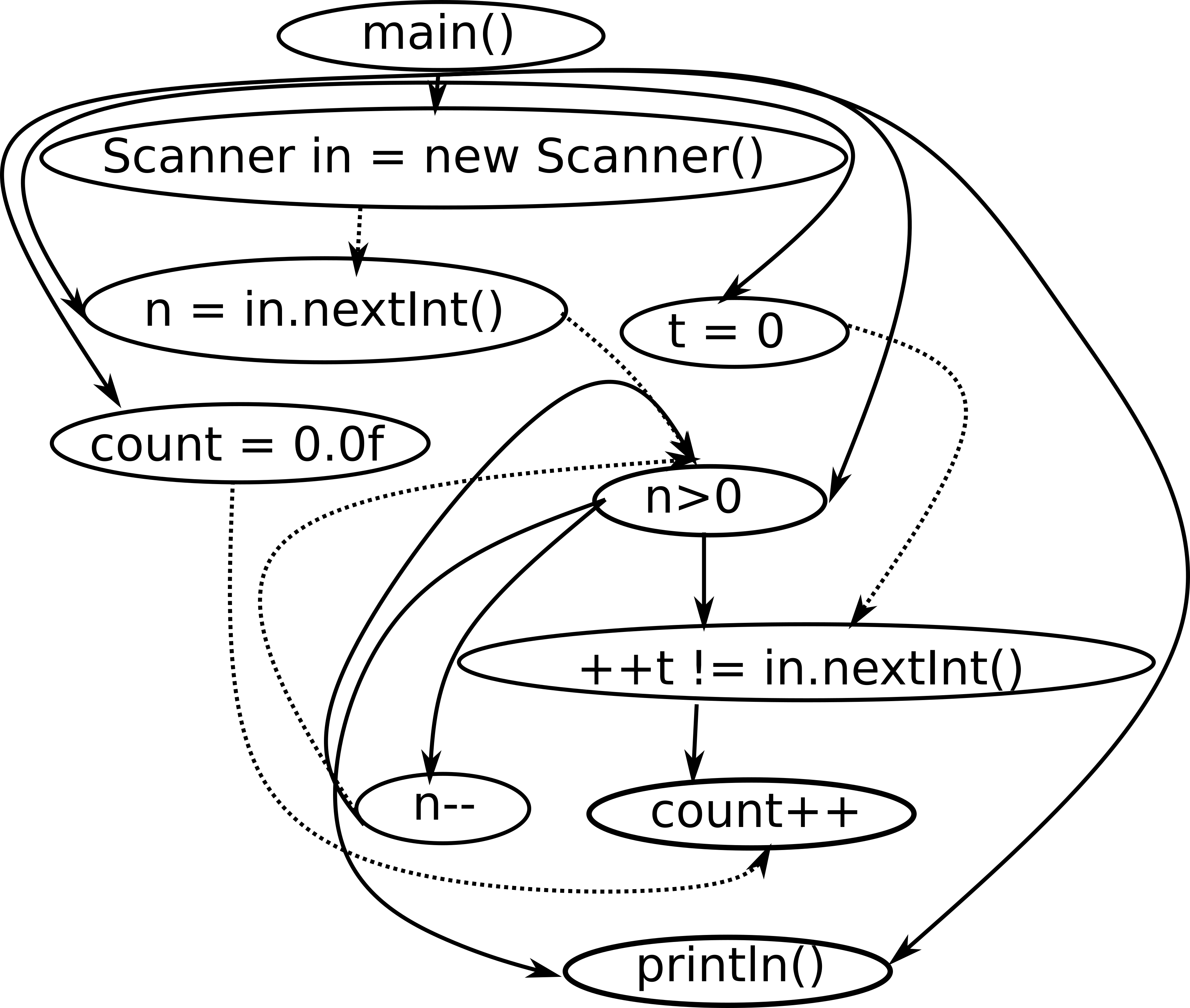}
              \caption{PDG for Listing \ref{code1}.}
        \label{PDG1}
    \end{minipage}
    \begin{minipage}{.5\textwidth}
        \centering
          \includegraphics[width=6.5cm,height=5.5cm]{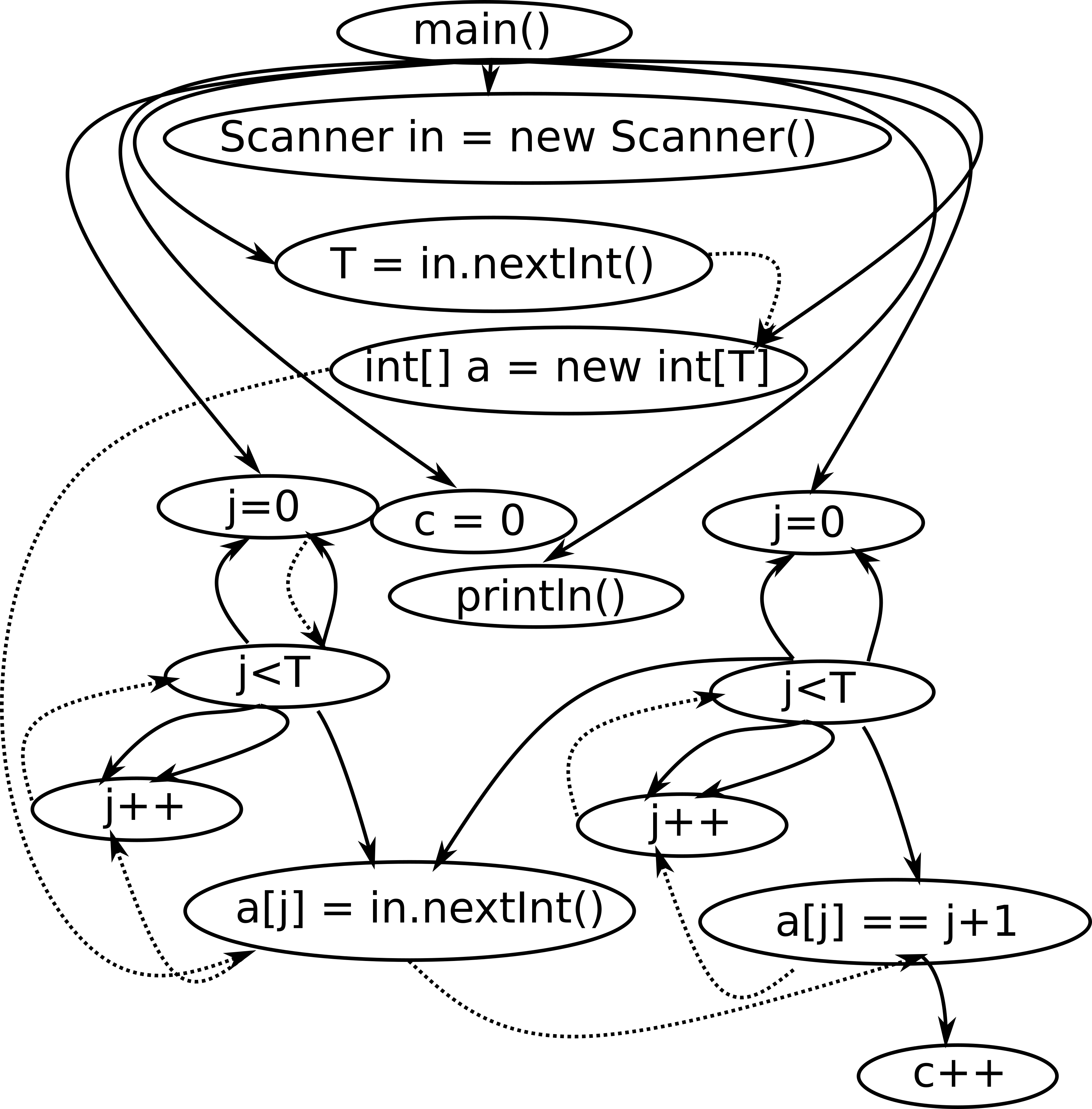}
              \caption{PDG for Listing \ref{code2}.}
        \label{PDG2}
    \end{minipage}
\end{figure*}
\par Addressing the above issues in this paper, we propose a new tool \tool, for measuring code functional similarity. \tool is based on two key insights. First, feature learning plays a significant role in measuring code similarity \cite{10.1145/2970276.2970326}. Thus, learned features should contain semantic information, specifically control and data dependence information, rather than structural information, such as lexical elements and high-level program constructs captured by ASTs. Code similarity based solely on syntactic features is too restrictive and also rigid in its expression compared to its more powerful and expressive notion based on program semantics or functionality. Hence, \tool uses the control and data dependence information from PDGs as a basis of similarity metrics. Second, to capture program semantics efficiently, one must capitalize on PDGs graphical structure. Therefore, \tool employs a graph-based deep neural network to learn program representation. Figure \ref{fig:approach overview} shows the overall architecture of the \tool.
\par We have implemented \tool in Java using the Soot optimization framework \cite{inproceedings} and Pytorch Geometric \cite{Fey/Lenssen/2019} deep learning library. We evaluated \tool on programming competition datasets and real-world datasets. Our empirical results show that \tool outperforms another state-of-the-art tool TBCCD \cite{10.1109/ICPC.2019.00021} and generalizes better on unseen 
code pairs.
\par We make the following contributions in this paper:
\begin{enumerate}
    \item 
    \textbf{A new code representation for semantic code clone detection}. To the best of our knowledge, our work is the first to learn code representation for code clone detection in two different manners: i) Using control and data dependence relations between the program statements to model code functional dependency ii) Treating control and data dependence edges differently to give respective importance to syntactic and semantic information while learning code representation for a code snippet.
   \item \textbf{A new code clone detection approach}. We proposed a new deep learning architecture for graph similarity learning. Our approach jointly learns the graph representation and graph matching function for computing graph similarity. In particular, we have used an attention-based siamese graph neural network to detect semantic clones. Our approach uses the control-dependent and data-dependent edges of PDGs to model the program's semantic and syntactic features. We have used attention to give higher weights to semantically relevant paths. The learned latent features are then used to measure code functional similarity.
    \item \textbf{A comprehensive comparative evaluation}. We developed a prototype tool \tool and evaluated it on popular benchmarks for code clones. Through a series of empirical evaluation, our results show that \tool outperforms the state-of-the-art-tool TBCCD. 
\end{enumerate}
\textbf{Paper Organization} Section \ref{sec:motivation} provides an example to motivate our approach. Section \ref{sec:background} presents an overview of basic concepts and code clone terminology. Section \ref{sec:approach} describes the code clone detection process and explains the Graph-based Siamese deep neural network used in our approach. Section \ref{sec:expts} details the experimental design and evaluation process. Section \ref{sec:results} discusses the results. Section \ref{sec:discussion} presents a qualitative analysis of \tool. Section \ref{threats to validity} discusses the threats to the validity of our proposed approach. Section \ref{sec:related work} surveys the related work, and finally, Section \ref{sec:conclusion} concludes our paper with a summary of findings.
\section{Motivation}
\label{sec:motivation}
\subsection{Motivating Example}
In this section, we present an example and our observations to motivate our approach.\newline
Listings \ref{code1} and \ref{code2} show two solutions submitted for the \textit{GoogleCodeJam} problem \textit{Goro Sort}. The problem involves an interesting method of sorting an array of natural numbers in which the array is shuffled $n$ times randomly to get it sorted.  The users have to report the minimum number of times shuffling is required to sort the array. 
\par Listing \ref{code1} implements the above functionality by first initializing an array of size $T$ with random numbers. It then checks if the current index element is equal to the index of the next element. The average number of hits required to sort the array was given by the size of the array ($T$) minus the number of times the element at the current index is equal to the next index.
\par Listing \ref{code2} implements the same functionality while taking input from the user at run time. It keeps the counter; if the current value is equal to counter$+1$, it reduces the average hits required by $-1$. Syntactically, Listings \ref{code1} and \ref{code2} are quite different. However, semantically, they are similar and will be classified as type 4 clones according to the taxonomy proposed in \cite{Roy07asurvey}.
\par The existing ML-based code clone detection approaches \cite{ijcai2018-394, 10.5555/3172077.3172312, 10.1109/ICPC.2019.00021} used syntactic and(or) lexical information to learn program features. For instance, TBCCD \cite{10.1109/ICPC.2019.00021} used tree-based convolution over abstract syntax trees (ASTs) to learn program representation. If we look at the ASTs of Listings \ref{code1} (for line $8-11$) and \ref{code2} (for line $5-8$) shown in Figures \ref{AST1} and \ref{AST2},
it's hard to infer that the two ASTs correspond to  similar programs. We executed TBCCD, trained on \textit{GoogleCodeJam} problems($2010-2017$), on this example and TBCCD caused a false negative by reporting Listings \ref{code1} and \ref{code2} as a non-clone pair. This led us to our first \textbf{Observation (O1):} \textit{To achieve accurate detection of semantic clones, we need to incorporate more semantic information while learning program representation.}
\par We then computed program dependency graphs (PDGs) for Listings \ref{code1} and \ref{code2}. The PDGs are shown in Figures \ref{PDG1} and \ref{PDG2}. If we look at the Figures \ref{PDG1} and \ref{PDG2}, we will observe that the flow of data and control between the two PDGs are similar, as PDGs approximate the semantic dependencies between the statements. However, PDGs suffer from scalability problems. The size of the PDGs can be considerably large. For a program with 40-50 lines of code, we can have around $100$ vertices and $100$ edges. This led us to our second \textbf{Observation (O2):} \textit{To learn important semantic features from the source code, a model should not weigh all paths equally. It should learn to give higher weights to semantically relevant paths.}
\par Source code is a complex web of interacting components such as classes, routines, program statements, etc. Understanding source code amounts to understanding the interactions between different components. Previous studies such as \cite{Binkley07sourcecode} have shown that graphical representation of source code is better suited to study and analyze these complex relationships between different components. Yet, the recent code clone detection approaches \cite{10.5555/3172077.3172312, 10.1145/3236024.3236068, 10.1145/2970276.2970326} do not make use of these well defined graphical structures while learning program representation. These approaches use deep learning models that do not take advantage of the available structured input, for example, capturing induced long range variable dependency between program statements.
This led us to our third \textbf{ Observation (O3):} \textit{To capitalize on the source code's structured semantic features, one might have to expose these semantics explicitly as a structured input to the neural network model.}
\subsection{Key Ideas}
Based on the above observations, we have created our approach with the following key ideas:
    \\
    $a)$ From observation 1, we learned program features from the PDG representation of source code to capture the program semantics. Such graphs enable us to capture the data and control dependence between the program statements.
    \\
    $b)$ From observation 2, we designed an attention-based deep neural network to model the relationship between the important nodes in the PDG. The attention-based model emphasizes learning the semantically relevant paths in the PDG necessary to measure code similarity.
    \\
    $c)$ From observation 3, we used a graph-based neural network model to learn the structured semantic features of the source code. We have encoded the source code's semantics and syntax into a graph-based structure and used a graph-based deep learning model to learn latent program features.

\section{Background}
\label{sec:background}
This section gives a brief overview of the basic concepts and defines the terminology used in the paper.
\subsection{Program Dependence Graphs}
Program Dependence Graph (PDG) is a directed attributed graph that explicitly encodes a program's control, and data dependence information \cite{PDG}. PDGs approximate program semantics. A node in a PDG represents a program statement such as an assignment statement, a declaration statement, or a method invocation statement, and the edges denote control or data dependence between program statements. 
\par A control dependence edge from statement $s_1$ to statement $s_2$ represents that $s_2$'s execution depends upon $s_1$. While data dependence edge between two statements $s_1$ and $s_2$ denotes that some component which is assigned at $s_1$ will be used in the execution of $s_2$. Control and data dependence relations in program dependence graphs are computed using control flow and data flow analysis. Formally control dependence can be stated as:
\par Given a control flow graph G for a program $P$, statements $s_1 \in G$ and $s_2 \in G$ are control dependent iff
\begin{itemize}
    \item there exists a directed path $\rho$ from $s_1$ to $s_2$ with any node $S$ in $P$ post-dominated ($S$ $\neq$ [$s_1$,$s_2$]) by $s_2$ and
    \item $s_1$ is not post-dominated by $s_2$
\end{itemize}
Data dependence can be formally defined as:
\par Two statements $s_1$ and $s_2$ are data dependent in a control flow graph if there exists a variable $v$ such that,
\begin{itemize}
    \item $v$ is assigned at statement $s_1$.
    \item $s_2$ uses the value of $v$.
    \item There exists a path between $s_1$ and $s_2$ along which there is no assignment made to variable $v$.
\end{itemize}
\par Program dependence graphs connect the computationally related parts of the program statements without enforcing the control sequence present in the control flow graphs \cite{PDG}. Hence, they are not affected by syntactical changes like statement reordering, variable renaming, etc. \cite{PDG}. These properties make program dependence graphs to be better representation to detect semantic clones. Horwitz \cite{Horwitz1990} and Podgurski and Clarke \cite{Podgurski1989} also showed that program dependence graphs provide a good representation to measure code semantic similarity.
\subsection{Concepts in Deep Learning}
\subsubsection{Artificial neural networks} ANNs \cite{10.1016/0004-3702(89)90049-0} or \textit{connectionist systems} are machine learning models that are inspired by the human brain. ANNs consist of several artificial neurons stacked together across several layers trained to discover patterns present in the input data.
\begin{figure*}[!tbp]
    \centering
    \includegraphics[width=\textwidth]{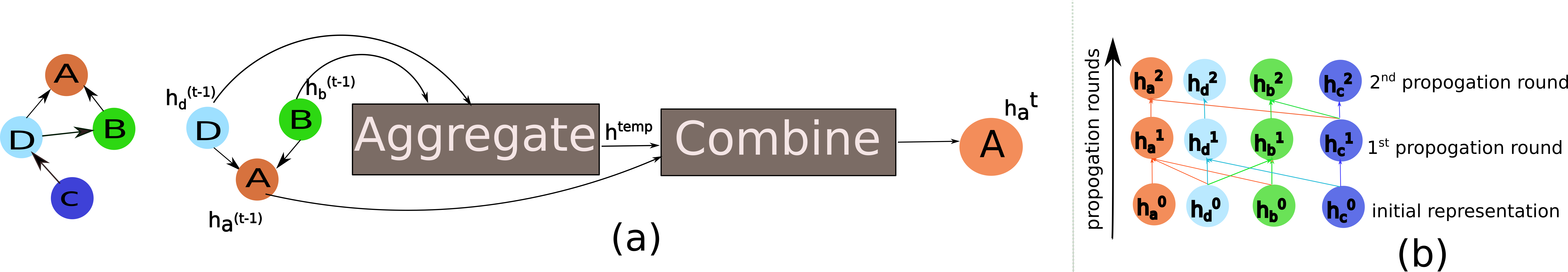}
    \caption{Visual illustration of the graph neural network framework. (a) Illustration of the $t^{th}$ layer of the graph neural network. The feature vectors $h_b^{t-1}, h_d^{t-1}$ from the neighbouring nodes of $A$ are \textit{aggregated} and \textit{combined} with $h_a^{t-1} $, the features of node $A$ from the $t-1^{th}$ layer. This constitutes the representation of node $A$ at the $t^{th}$ layer. (b) Illustration showing multiple rounds of propagation in a graph neural network. At the $n^{th}$ propagation round, a node receives information from each of its neighbors that are $n$ hops away. For example, node A at propagation round 1 receives messages from its one-hop neighbors D and B. At propagation round two, it receives information from its two-hop neighbor, i.e., node C, and so on.}
    \label{fig:gnnframework}
\end{figure*}
\subsubsection{Deep Learning} Deep Learning covers a set of algorithms that extracts high-level representations from the input data. Deep learning models use artificial neural networks with several layers of neurons stacked together. Each layer learns to transform the previous layer's output into a slightly more abstract representation of the input data. Deep neural networks can readily model the linear and complex, non-linear relationships between input data and the desired output prediction. Many variants of Deep Neural Networks (DNN) exist, such as, recurrent neural networks \cite{lstm}, convolutional networks \cite{article}, graph-based neural networks \cite{bronstein2016geometric} etc. In this work, we make use of graph-based neural networks.
\subsubsection{Graph Neural Networks}
\par DNNs have shown unprecedented performance in many complex tasks such as image processing \cite{hu2018squeeze} and neural machine translation \cite{DBLP:journals/corr/abs-1804-07755}. DNN architectures like transformers \cite{vaswani2017attention} and convolutional networks \cite{DBLP:journals/corr/abs-1709-01507} have often demonstrated performance at par with humans. The key reason behind the success of DNNs is the model's ability to take input data directly and learn to extract feature representations relevant to a complex downstream task like classification or retrieval. 
\par Despite state-of-the-art results, the above models do not perform well in non-Euclidean domains such as graphs and manifolds. The inherent complexity of the data, variegated structural and topological information hampers the ability to gain true insights about the underlying graphical data \cite{bronstein2016geometric,Zhangsi,zonghan}. Nevertheless, one may have to deal with graph-structured data in various fields.
For example, in software engineering, programs are modeled as graphs (ASTs, PDGs, etc.) for automatic code summariation \cite{leclair2020improved}, identifying vulnerabilities \cite{DBLP:conf/icait/WangZWXH18}, and bug-fixing activities \cite{8811910}.
\par Dealing with non-Euclidean structured data implies that there are no such properties as the shift-invariance and the vector space structures \cite{bronsteinBLSV16}. Hence, convolutions and filterings are not well defined here. Therefore, spectral-domain \cite{bruna2013spectral} and spatial domain \cite{4773279} techniques have been adopted to learn representation of the graph-structured data.
\par Our work makes use of the technique from the spatial (vertex) domain. Spatial graph convolutions define convolution operations based on the node's spatial connections and are built on the idea of message passing. The graph convolutional operator learns a function $f$ to generate node $v_{i}$'s representation by aggregating its own features $h_{i}$ and neighbor's features $h_{j}$. Multiple iterations of graph convolution are performed to explore the depth and breadth of the node's influence. Each iteration uses node representation learned from the previous iteration to get the representation for the current one. For instance, in the first iteration of graph convolution, information flow will be between first-order neighbors; in the second iteration, nodes will receive information from second-order neighbors, i.e., neighbor's neighbor. Thus traversing this way, after multiple iterations, each node's final hidden representation will have information from a further neighborhood. Figure \ref{fig:gnnframework} depicts the general framework for spatial graph convolutions.
\subsubsection{Siamese Neural Networks}
Siamese neural network or twin network \cite{siamese, siamese2} is an artificial neural network for similarity learning that contains two or more identical sub-networks sharing the same set of weights and parameters. The Siamese neural networks are trained to learn the similarity between the input data. They try to learn a mapping function such that the distance measure between the learned latent features in the target space represents the semantic similarity in the input space.
\subsection{Representation learning in software engineering}
Treating program as data objects and learning syntactically and semantically meaningful representations have drawn a great deal of interest \cite{aloncode2seq,aloncode2vec,allamanisprogramgraphs}.
\begin{lstlisting}[float=t,caption=Different clone types of gcd,captionpos=b,label=exampleclones,firstnumber=1,frame=tlrb]
    //orginal code snippet
    static int gcd(int a, int b) 
    { 
        if (b == 0) 
            return a; 
        return gcd(b, a % b);  
    } 
     
    //type 1 clone
    static int gcd1(int a, int b) { 
        if (b == 0){ 
            return a; 
        }
        return gcd1(b, a % b);  
    } 
    
    //type 2 clone
    public static int gcd2(int no1, int no2) {
	    if (no2 == 0) {
	        return 1;
    	} 
	    return gcd2(no2, no1 % no2);    
	}
    
    //type 3 clone
    public static int gcd3(int m, int n) {
        if (0 == n) {
            return m;
        } else {
            return gcd3(n, m % n);
        }
    }
    
    //type 4 clone
    static int gcd4(int a, int b) {
        while (b != 0) {
            int t = b;
            b = a % b;
            a = t;
        }
        return a;
    }
\end{lstlisting}

Following the success of deep neural networks in natural language processing, computer vision, etc., learning tasks on source code data have been considered recently. Program synthesis \cite{Chen2020Neuralprogramsynthesis, Programsynthesis2018shinrichard}, program repair \cite{wangneuralprogramembedding}, bug localization \cite{LiYiWangOOPSLA,allamanisprogramgraphs}, and source code summarization \cite{MiltiadisAllamaniscodesummarisation} are some of the well-explored areas. The idea is to use the knowledge from the existing code repositories to enable a wide array of program analysis and maintenance tasks. The key step is to design a precise and semantically meaningful program representation that neural networks will use in the array of downstream tasks.  
\par Most existing approaches use two kinds of program representations extracted from static and dynamic program analysis techniques. These representations can further be categorized into syntactic and semantic program representations. Abstract Syntax Trees (ASTs), Control Flow Graphs (CFGs), Call Graphs, etc., represent the program's syntactic structure while Program Dependence Graphs(PDGs), execution traces, etc., capture program semantics. These representations help to transform programs in an appropriate form to deep learning models. 
\subsection{Deep learning for code clone detection}
\par Learning-based techniques automatically (using neural networks) learn a continuous-valued feature vector representing program semantics and syntax to learn similarities between code snippets. This feature vector is then compared directly (using a distance-based classifier) or is passed to a neural network classifier to predict similarity. 
\par For instance, White et al. \cite{10.1145/2970276.2970326} used a recursive neural network to learn program representation. They represented source code as a stream of identifiers and literals and used it as an input to their deep learning model. Tufano et al. \cite{8595238} used a similar encoding approach as \cite{10.1145/2970276.2970326} and encoded four different program representations- identifiers, Abstract Syntax Trees, Control Flow Graphs, and Bytecode. They then used a deep learning model to measure code similarity based on their multiple learned representations. Zhao and Huang \cite{10.1145/3236024.3236068} used a feed-forward neural network to learn a semantic feature matrix constructed from a program's control flow and data flow graphs. Yu et al. \cite{10.1109/ICPC.2019.00021} used a tree-based convolutional neural network to detect code clones.
\par These code clone detection approaches have used syntactic and lexical features to measure code similarity. They do not exploit the source code's available structured semantics, even though this information might be useful to measure code functional similarity. Hence to overcome the limitations of existing approaches, we have proposed a novel code clone detection tool \tool. \tool uses PDGs and graph-based neural networks to learn structured semantics of the source code. Section \ref{sec:approach} explains the code clone detection process of \tool.
\subsection{Terminologies}
This paper follows the well-accepted definition and terminologies from \cite{Roy07asurvey}:\newline
\textbf{Code Fragment}: A continuous segment of a code fragment is denoted by a triplet $\langle c,s,e \rangle$, where $s$ and $e$ are start and end lines respectively, and $c$ is the code fragment. \newline
\textbf{Code clones} are pairs of similar code snippets existing in a source file or a software system. Researchers have broadly classified clones into four categories stretching from syntactic to semantic similarity \cite{Roy07asurvey}:
\begin{itemize}
    \item \textbf{Type-1 clones (textual similarity)}: Duplicate code snippets, except for variations in white space, comments, and layout.
    \item \textbf{Type-2 clones (lexical similarity)}: Syntactically identical code snippets, except for variations in the variable name, literal values, white space, formatting, and comments.
    \item  \textbf{Type-3 clones (syntactic similarity)}: Syntactically similar code snippets that differ at the statement level. Code snippets have statements added, modified, or deleted w.r.t. to each other.
    \item \textbf{Type-4 clones (semantic similarity)}: Syntactically different code snippets implementing the same functionality.
\end{itemize}
\begin{figure*}
    \centering
    \includegraphics[width=\textwidth]{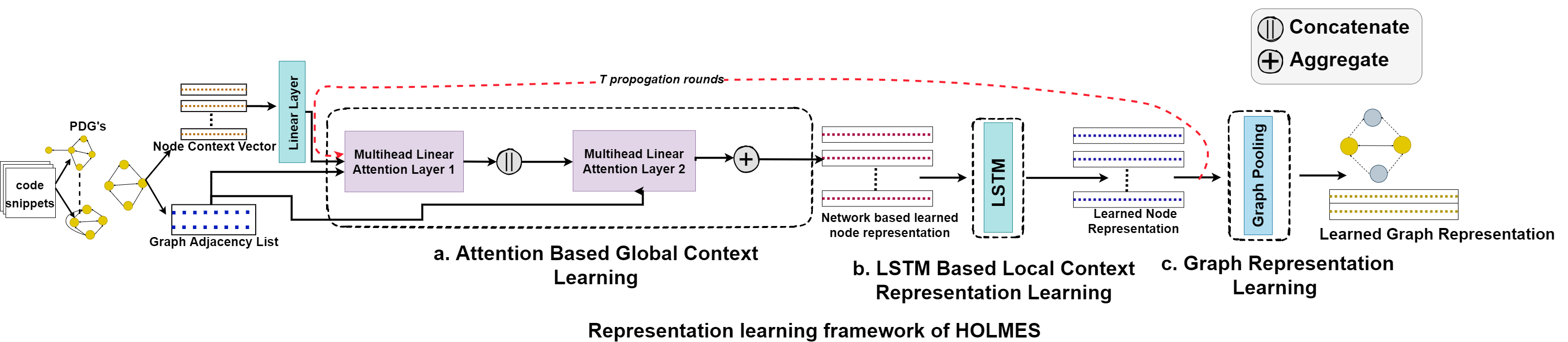}
    \caption{The architecture of one branch of the Siamese neural network is shown in Figure \ref{fig:approach overview}. (a) Our model first parses the given Java methods in the datasets to build PDGs. Node feature matrix and graph adjacency matrix are extracted from the source code. \tool then passes this as input to a multi-head masked linear attention module, which learns the importance of different sized neighborhood for a node. (b)The attention module outputs the set of learned node features that are then passed through an LSTM, which extracts and filters the features aggregated from different hop neighbors. (c)The learned node features are then passed to a graph pooling module. Graph pooling employs a soft attention mechanism to downsample the nodes and to generate a coarsened graph representation.}
    \label{fig:Approach}
\end{figure*}
Listing \ref{exampleclones} enumerates different clone types from the BigCloneBench dataset. The original code snippet (starting from line 2) computes the greatest common divisor (gcd) of two numbers. The Type-1 clone (starting from line $10$) of the original code snippet is identical except for the formatting variation. The Type-2 clone (starting from line $18$) have different identifier names (no1 and no2). Type-3 clone (starting from line $26$) of the original code snippet is syntactically similar but differs at the statement level. Finally, the Type-4 clone of the original code snippet computes gcd using a completely different algorithm. There exists no
syntactical similarity between the original snippet and its Type-4 clone.


\section{Our Approach}
\label{sec:approach}
This section discusses the details of our graph neural network architecture that is used to learn high level program features from program dependency graphs.
Figure \ref{fig:Approach} shows an overview of one branch of the Siamese neural network shown in Figure \ref{fig:approach overview} . The following subsections give details of the main steps of the proposed approach.
\subsection{Attention Based Global Context Learning}
\label{subsec:gat}
Our work builds on Graph Attention Networks (GAT) \cite{GAT}, and we summarize them here. Given a program dependence graph $G = (V,E,A,X)$, we have a set of $V$ vertices representing program statements, and a list of directed control and data-dependent edges $E = (E_{1},E_{2})$. $A$ denotes the adjacency matrix of $G$, where $ A \in \mathbb{R} ^{|V| \times |V|}$ with $A_{ij}$ = 1 if $e_{ij} \in E$, and $A_{ij}$ = 0 if $e_{ij} \notin E$. The node feature matrix is represented by $ X\in \mathbb{R}^{|V| \times d} $ with $ x_{v} \in \mathbb{R} ^{d} $ denoting the feature vector of vertex $v$.
\par For every node $v \in V$, we associate a feature vector $x_{v}$, representing the type of statement it belongs to. We considered the following $18$ types: \texttt{Identity, Assignment, Abstract, Abstract Definition, Breakpoint, Enter Monitor, Exit Monitor, Goto, If, Invoke, LookupSwitch, Nop, Return, Return void, Throw, JTableSwitch} corresponding to the types of the statements used by Soot's internal representation. We encode this statement type information into an $18$-dimensional one-hot encoded feature vector. For example, the statement $x = y + z$ is of type \texttt{Assignment statement} and will be represented as \textbf{[0,1,0,0,0,0,0,0,0,0,0,0,0,0,0,0,0,0]}. In the first place, to obtain initial node vectors, we pass node features through a linear transformation layer:
\begin{equation}\label{eq:mlp}
    H^{0} = X \times W + b
\end{equation}
Where $W$ and $b$ are the learnable weights and bias of the linear layer. $H^{0} \in \mathbb{R}^{|V| \times d^{'}}$ denotes the initial node embeddings matrix with $h^{0}_{i}$ representing embedding vector for a single node $i \in V$. Line \ref{algo:linlayer} in Algorithm \ref{algo} inside function $ComputeGFeatures$ denotes the above action.

\begin{algorithm}[th]
\caption{Code Similarity Detection}
\label{algo}
\SetKwInput{KwInput}{Input}
\SetKwInput{KwOutput}{Output}
\DontPrintSemicolon
\KwInput{$T$ rounds of propagation, $train\_Data$} 
\KwOutput{$code\_similarity$} 
  \SetKwProg{Fn}{Define}{:}{\KwRet}
  \Fn{ComputeNodeFeatures($H^{0}$, $A$, $T$)}{
  $H_{F} = [H^{0}]$\;
  \tcp{$C^{0}$ is the initial cell state of $LSTM$ }
  \textbf{Initialize} \textbf{Array } $C^{0}$\;
 \For{$t\gets1$ \KwTo $T$ }{
    \tcp{Attention block 1 (Eqn. (\ref{eq:a1}))}
    $H^{'}$ = $Attn_{1}\left(A,H^{(t-1)};\phi_1\right)$ \label{A1:CNF:a1}\;
    \tcp{Attention block 2 (Eqn. (\ref{eq:a2}))}
    $H^{''(t-1)} = Attn_2\left(A,H^{'};\phi_2\right)$\label{A1:CNF:a2}\;
    \tcp{$t^{th}$ propagation round (Eqn. (\ref{eq:GPdepth}))}
    $H^{(t)}, C^{(t)}$ = $LSTM(H^{''(t-1)},C^{(t-1)})$\label{A1:CNF:lstm}\;
    \tcp{Final node features (Eqn. (\ref{eq:cat_node_reps}))}
    $H_{F} = CONCAT(H^{(t)})$\label{A1:CNF:cat_features}\;
    }
    
    \KwRet $H_{F}$\;
    }
  \;
   \SetKwProg{Fn}{Define}{:}{\KwRet}
  \Fn{GraphPooling($H_{node}$)}{
    $H_{G} = a(MLP(H_{node})) \odot MLP(H_{node})$\;
    \KwRet $H_{G}$\;
    }
    \;
  \SetKwProg{Fn}{Define}{:}{\KwRet}
  \Fn{ComputeGFeatures($X$, $A_{c}$, $A_{d}$, $T$)}{
        $H^{0} = X \times W + b$\; \label{algo:linlayer}
        $H_{d} = $ComputeNodeFeatures$(H^{0}, A_{d}, T)$\;
        $H_{c} = $ComputeNodeFeatures$(H^{0}, A_{c}, T)$\;
        $H_{final} = H_{d} + H_{c}$\;
        $G_{final} = $GraphPooling$(H_{final})$\;
        \KwRet $G_{final}$\;
    }
    \;
\SetKwProg{Fn}{Define}{:}{\KwRet}
  \Fn{Train($train\_data$, $T$)}{
  \While {$not\_converged$} {
    \While {$data$ in $train\_data$} {
        $X_{1} = data.X_{1}$\;
        $X_{2} = data.X_{2}$\;
        $A_{c1} = data.A_{control1}$\;
        $A_{d1} = data.A_{data1}$\;
        $A_{c2} = data.A_{control2}$\;
        $A_{d2} = data.A_{data2}$\;
        $Y = data.Y$\;
        $G_{1} = $ComputeGFeatures$(X_{1}, A_{c1}, A_{d1},T)$\;
        $G_{2} = $ComputeGFeatures$(X_{2}, A_{c2}, A_{d2},T)$\;
        $featureRep = CONCAT(G_{1},G_{2})$\;
        $featureRep = a(MLP(featureRep))$\;
        $similarity = a(MLP(featureRep))$\;
        $loss = LOSS(similarity,Y)$\;
        $Update\_Optimizer(loss)$\;
    }
    }
    }
\end{algorithm}

Next, to obtain node features for a given graph, we learn an adaptive function $\varphi (A, H; \phi) $ parametrized by $\phi$ similar to  GAT \cite{GAT}.
The input to the function is the set of node features $\{h^{0}_{1},h^{0}_{2}, \cdots ,h^{0}_{|V|}\}$ obtained from Equation \eqref{eq:mlp}. The function $\varphi$
then outputs the set of new node features  $\{h^{'}_{1},h^{'}_{2} ,\cdots ,h^{'}_{|V|}\}$ as the output of the first attention block. It computes the self-attention on nodes based on the graph structural information where a node $v_{j}$ attends to its one-hop neighboring node $v_{i}$, i.e., if $(v_{i},v_{j}) \in E$. The attention mechanism $a : \mathbb{R}^{d^{'}} \times  \mathbb{R}^{d^{'}} \rightarrow \mathbb{R}$ computes attention coefficients
\begin{equation}\label{eq:node_imp}
    e_{ij} = a(Wh_{i},Wh_{j})
\end{equation}
Equation \eqref{eq:node_imp} denotes the importance of node $j's$ features for node $i$. {\color{blue}The scores are then normalized using the softmax function}
\begin{equation}\label{eq:scores}
     \alpha_{ij} = \frac{exp(e_{ij})}{\sum_{j \in N(i)} exp(e_{ij})}
\end{equation}
The attention scores computed in Equation \eqref{eq:scores} are then used to output a linear combination of features of node  $v_{j}, ~\forall j \in N(i)$ that will be used as the final output features of node $v_{i}$.
\par We have used two attention modules to learn node representation. The output of attention module $1$ with eight different attention heads is shown by Equation \ref{eq:a1}
\begin{equation}\label{eq:a1}
    h_{i}^{'} = \|_{k = 1}^{8} \sigma \left( \sum_{j \in N(i)} \alpha_{ij}^{k}W^{'k}h_{j}^{0} \right)
\end{equation}
Where $h_{i}^{'}$ denotes the intermediate representation after the first attention block, $\alpha$ denotes the corresponding attention scores, $\sigma$ is the sigmoid activation function, $W'$ are the weight parameters in the first attention block, and  $\|$ represents the concatenation of the attention coefficients from eight different heads. Equation \ref{eq:a2} represents the output of attention module $2$. Here, we have aggregated the output from different attention nodes.
\begin{equation}\label{eq:a2}
    h_{i}^{''} = \sigma \left( \sum _{k=1}^{6} \sum_{j \in N(i)} \beta_{ij}^{k}W^{''k}h_{j}^{'} \right)
\end{equation}
 Here, $\beta$ represents the attention scores computed for attention module $2$, $W''$ are the weight parameters in the second attention block, and $h_{i}^{''}$ represents the learned node features at the output of the second attention block. This node representation $h_{i}^{''}$ is input to the LSTM module as $h_{i}^{''(0)}$ in the first propagation round. As shown in Figure \ref{fig:approach overview}, the LSTM module's output, $h_{i}^{(1)}$, is fed back to the first attention block as an input. The process described above is repeated to update the node representations, with the output of the second attention block $h_{i}^{''(t-1)}$ as input to the LSTM at the $t^{th}$ propagation round. The $T$ propagation rounds result in learned representations of the individual nodes that capture the semantic context through the graph's structural information and the learnable attention-based weights. Lines \ref{A1:CNF:a1}-\ref{A1:CNF:lstm} of Function $ComputeNodeFeatures$ defined in Algorithm \ref{algo} presents the above exposition. The parameter sets $\phi_1$ and $\phi_2$ comprise all the first and second attention blocks' parameters, respectively.
\subsection{LSTM Based Local context Learning}
\label{subsec:lstm}
The multi-head attention mechanism enables node representations to capture the context from their one-hop neighbors. However, the semantic context within a code fragment typically requires a broader context provided by a node's $t$-hop neighbors. Inspired by the architecture of the adaptive path layer in GeniePath \cite{GeniePath}, we use an LSTM layer, which when combined with the multi-head attention previously described, helps learn node representations that are better equipped to capture the semantic information of the code fragment.
The input to the LSTM model at the $(t+1)^{th}$ propagation round is $h_i^{''(t)}$, the representation of the $i^{th}$ node. This strategy allows the node representations to capture the context from its $t$-hop neighbors \cite{GeniePath}.\\
We initialize the LSTM cell with random  values. The cell state $C^{(t)}_j$ corresponding to the $j^{th}$ node ($j\in V$) is updated in the $t^{th}$ propagation round, effectively aggregating information from $t$-hop neighbors of node $j$. The node representation is then accordingly updated as a function of the cell state. The update Equations are presented below.
\begin{equation}\label{eq:GPdepth}
\left.\begin{aligned}
& i_{j} = \sigma(W_{i}h^{''(t-1)}_{j}), &f_{j} = \sigma(W_{f}h^{''(t-1)}_{j}),\\  
& o_{j} = \sigma(W_{o}h^{''(t-1)}_{j}), &\widetilde{C}_j\! =\! tanh(W_{c} h_{j}^{''(t-1)}),\\
& C_{j}^{(t)}\! =\! f_{j}\! \odot\! C_{j}^{(t-1)}\! +\! i_{j}\!\odot\! \widetilde{C}_j,& h_{j}^{(t)} = o_{j} \odot tanh(C_{j}^{(t)})
    \end{aligned}
\right\}
\end{equation}
Where $\odot$ denotes element-wise multiplication. The input gate of LSTM $i_{j}$ is being used to extract new messages from the input $h_{j}^{''(t-1)}$ and are added to memory $C_{j}^{(t)}$. The gated unit $f_{j}$ is the forget gate used to filter out unwanted messages from the old memory $C_{j}^{(t-1)}$. Lastly, the output gate $o_{j}$ and the updated memory $C_{j}^{(t)}$ are used for constructing the final node representation $h_{j}^{(t)}$ at $(t)^{th}$ propagation round for node $j$.
\begin{figure}
    \centering
    \includegraphics[height=3.5cm,width=3.5cm]{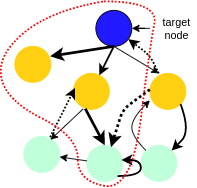}
    \caption{A synthetic example showing explored receptive path (area covered by the dotted red line) for the target node. The edge thickness denotes the received attention scores while learning features for the target node. Control and data-dependent edges are shown through solid and dotted edges, respectively.}
    \label{fig:lstm}
\end{figure}

\par Figure \ref{fig:lstm} conveys the above exposition through a synthetic example. \tool tries to filter and aggregate meaningful features from different two-hop neighbors while learning representation for the target node. The multi-head attention module in each propagation round attends to different neighbors (edge width in Figure \ref{fig:lstm} denotes the importance of different hop neighbors at each propagation step). The LSTM module filters and aggregates the messages received from different hop neighbors over multiple propagation rounds. The area covered by the red dotted line denotes the relevant neighboring nodes (receptive path) for learning the feature representation of the target node.
\subsection{Graph Representation Learning with Jumping Knowledge Networks}
\label{subsec:gp}
\begin{figure}[t]
    \centering
    \includegraphics[width=0.5\textwidth]{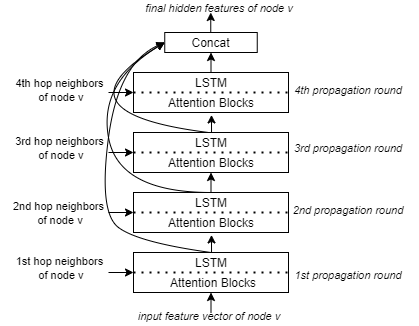}
    \caption{Illustration of the $4$-layer architecture of \tool using Jumping Knowledge Networks (JK nets) defined by Xu et al. \cite{JK}. At every propagation round t, the feature vector of node $v$ is aggregated with its $t^{th}$ order neighbors. At the last layer ($4^{th}$ propagation round), all the hidden feature vectors from all the propagation rounds are concatenated to constitute the final hidden representation for node $v$. The concatenation of hidden features from all the propagation rounds ensures that the features learned from $n^{th}$ hop neighbors during different propagation rounds are retained till last and also reflected in the final hidden representation of node $v$.}
    \label{fig:JK}
\end{figure}
To learn the high-level program features from the program dependence graphs, our graph neural network (GNN) model iteratively aggregates the node features from different $n^{th}$ hop neighbors via message passing scheme, described in Sections \ref{subsec:gat} and \ref{subsec:lstm}. 
\par To learn the diverse and locally varying graph structure and the relations between program statements effectively, a broader context is needed, i.e., the GNN model should explore the deeper neighborhood. We observed that, while aggregating features from the different $n^{th}$ hop neighbors (going till depth $4$ -- i.e. $n \in {1,2,3,4}$), our GNN model's performance degrades (also shown in Figure \ref{fig:jkgapexpt}). 
\par Thus, to stabilize the training and learn the diverse local neighborhood for each node, we employed Jumping Knowledge Networks (JK) \cite{JK}, shown in Figure \ref{fig:JK}. JK combines (concatenates; denoted by $\|$ operator below) the hidden features ($H^{(t)}$ defined on Line \ref{A1:CNF:lstm}, Algorithm \ref{algo}) learned at each GNN iteration independently:
\begin{equation}\label{eq:cat_node_reps}
    H_F = \left[ H^{0} \| H^{(1)} \| \cdots \| H^{(T)} \right]
\end{equation}
Line  \ref{A1:CNF:cat_features} of Function $ComputeNodeFeatures$ defined in Algorithm \ref{algo} conveys the above description.

\par For obtaining graph level representation from the learned node feature vectors, we have employed a soft attention mechanism proposed by Li et al. \cite{ggnn}:
\begin{equation}
    H_{G} =  \left(\sum_{i \in V}\sigma \big(MLP(h_{i}^{(T)})\big) \odot MLP(h_{i}^{(T)}) \right) 
\end{equation}
where $T$ denotes the $T$ rounds of propagation and $\sigma \left(MLP(h_{i}^{(T)})\right)$ computes the attention scores. The attention scores act as a filtering mechanism that helps to pull out irrelevant information. The Function $GraphPooling$ defined in Algorithm \ref{algo} shows the above exposition.

\subsection{Edge-attributed PDGs} 
PDGs use data dependence and control dependence edges to capture the syntactic and semantic relationships between different program statements. Control dependence edges encode program structure while data dependence edges encode the semantics.
\par Hence, to leverage the available syntactic and semantic information more effectively, we propose to learn program representations corresponding to each edge type. Therefore, given a program dependence graph $G$, we will learn two separate node feature matrix $H_{data}$ and $H_{control}$. $H_{data}$ represents the learned node feature matrix corresponding to the subgraph of $G$ induced by data dependence edges, and $H_{control}$ represents the learned node feature matrix corresponding to the subgraph induced by control dependence edges in $G$. Next, to obtain the final representation for the nodes in $G$, we do the vertex wise addition of $H_{data}$ and $H_{control}$. \newline
\begin{equation}\label{EdgeD}
\left.\begin{aligned}
    H_{data} = ComputeNodeFeatures (H_{0},A_{d},T)\\
     H_{control} = ComputeNodeFeatures (H_{0},A_{c},T)\\
     H_{final} = H_{data} + H_{control}
     \end{aligned}
\right\}
\end{equation}
Thereafter, to obtain the final graph representation $G_{final}$ we applied graph pooling defined in Section \ref{subsec:gp} on the learned node feature matrix $H_{final}$.
\begin{equation}\label{finalG}
    G_{final} = GraphPooling(H_{final})
\end{equation}

\subsection{Implementation and Comparative Evaluation}
\label{subsec:parmeter}
We have used the Soot optimization framework \cite{inproceedings} to build program dependence graphs. To compute the control dependence graph, we first build a control flow graph. Then Cytron's method \cite{cdg} is used to compute control dependence. For computing data dependence graph, reaching definition \cite{RDAnalysis} and upward exposed analysis \cite{Upwardexpose} is used.
\par We have used the PyTorch geometric \cite{torchgeometric} deep learning library to implement \tool. All LSTMs have a single LSTM layer with $100$ hidden units. We have used the LeakyReLU \cite{LReLU} as the non-linear activation function with a negative slope of $0.02$ and a sigmoid layer at the output for classification.
The network is initialized using the Kaiming Uniform method \cite{kaiming}. The Siamese network is trained to minimize binary cross-entropy loss given in Equation \ref{bce} using Adam \cite{kingma2014adam} optimizer with a learning rate set to $0.0002$ and batch size to $50$.
 \begin{equation}\label{bce}
        BCELoss = -\frac{1}{N}\sum_{i=1}^{N}y_{i}\times log(p(y_{i})) + (1-y_{i})\times log(1-p(y_{i}))
    \end{equation}
 Where $y_{i}$ denotes the true binary label, and $p(y_{i})$ denotes predicted probability (similarity score). $N$ is the number of samples in the dataset. The output of Algorithm \ref{algo} is the similarity score. To determine the decision threshold ($\epsilon$), we employed a threshold moving approach. We first predicted the probability for each sample on the validation set and then converted the probabilities into the class label by varying $\epsilon$ from $[0.2-0.8]$ with the step size of $0.1$. We evaluated the class labels on each threshold value in the range and selected $\epsilon$ on which we got the maximum F1-score on the validation set. This threshold was then used to evaluate the samples in the test set and is also used in the experiments defined in \ref{rq:rq2}
 \par For the BigCloneBench (BCB) dataset since it does not provide the input files' dependency information, we used JCoffee \cite{Jcoffee} to infer missing dependencies to generate PDGs. JCoffee infers missing dependencies based on the compiler's feedback in an iterative process. With JCoffee, we successfully compiled $90\%$ of the snippets from the BCB dataset.
  We compared \tool with the state-of-the-art code clone detection tool TBCCD \cite{10.1109/ICPC.2019.00021}. Other recent machine-learning-based code clone detection tools namely, CDLH \cite{10.5555/3172077.3172312} and CDPU \cite{ijcai2018-394} do not have their implementation available open-source. 
  DeepSim \cite{10.1145/3236024.3236068} does not provide implementation details of the semantic feature matrix construction. Thus, we could not replicate their experimental settings and hence do not perform a comparative evaluation with these approaches. Moreover, we did not compare Holmes with Deckard\cite{4222572}, RtvNN \cite{10.1145/3236024.3236068} and Sourcerer \cite{7886988} as TBCCD significantly outperformed these approaches. Therefore, TBCCD became our natural choice for comparative evaluation.

\section{Experimental Design}
\label{sec:expts}
This section details the comprehensive evaluation of \tool.  
Specifically, we aim to answer the following research questions (RQs):\newline
\textbf{RQ1}: How effective is \tool as compared to other state-of-the-art approaches? \newline
\textbf{RQ2}: How well \tool generalizes on unseen projects and data sets?
\subsection{Dataset Collection}
\begin{table}[t]
  \caption{Dataset Statistics.}
  \label{tab:statistics}
   \def\arraystretch{1.5}
  \begin{tabular}{|M{0.05\textwidth}|M{0.08\textwidth}|M{0.11\textwidth}|M{0.05\textwidth}|M{0.08\textwidth}|}
    \hline
    \textbf{Dataset} & \textbf{ Language} & \textbf{Project Files} & \textbf{ Clone Pairs} & \textbf{Non-clone Pairs}\\
    \hline
    GCJ & Java & $9,436$ & $4,40,897$ & $5,00,000$\\
    \hline
    SeSaMe & Java & $11$ Java projects & $93$ & n.a.\\
    \hline
    BCB & Java & $9134$ & $6,50,000$ & $6,50,000$\\
    \hline
  \end{tabular}
\end{table}
\begin{table}
    \centering
     \caption{Percentage of clone-types in BigCloneBench.}
    \label{tab:Percentage of clone-types in BigCloneBench}
    \def\arraystretch{1.5}
    \begin{tabular}{|c|c|c|c|c|c|c|}
    \hline
        \textbf{Clone Type }& T1 & T2 & ST3 & MT3 & WT3/T4\\
        \hline
        \textbf{Percentage($\%$)} &  $0.005$ & $0.001$ & $0.002$ & $0.010$ & $0.982$ \\
        \hline
    \end{tabular}
\end{table}
Our experiments make use of the following datasets to evaluate the effectiveness of the proposed approach:
    \\\\
    $\textbf{1)}$ \textbf{Programming Competition Dataset}: We followed the recent work \cite{10.1145/3236024.3236068} and used code submissions from GoogleCodeJam\footnote{https://code.google.com/codejam/past-contests} (GCJ). GCJ is an annual programming competition hosted by Google. GCJ provides several programming problems that participants solve. The participants then submit their solutions to Google for testing. The solutions that pass all the test cases are published online. Each competition consists of several rounds. 
    \par However, unlike the recent work \cite{10.1145/3236024.3236068} that used $12$ different functionalities in their experiments, we collected $9436$ solutions from $100$ different functionalities from GCJ. Thus, building a large and representative dataset for evaluation. Detailed statistics are reported in Table \ref{tab:statistics}. Programmers implement solutions to each problem, and Google verifies the correctness of each submitted solution. All $100$ problems are different, and solutions for the same problems are functionally similar (i.e., belonging to Type 3 and Type 4 clone category) while for different problems, they are dissimilar.
    \\\\
    $\textbf{2)}$ \textbf{Open Source Projects}: We experimented with several open-source real-world projects to show the effectiveness of \tool's learned representations.
    
    $\textbf{a)}$ \textbf{SeSaMe dataset}. SeSaMe \cite{10.1109/MSR.2019.00079} dataset consists of semantically similar method pairs mined from $11$ open-source Java repositories. The authors applied text similarity measures on Javadoc comments mined from these open source projects. The results were then manually inspected and evaluated. This dataset reports $857$ manually classified pairs validated by eight judges. The pairs were distributed in a way that three judges evaluated each pair. The authors have reported semantic similarity between pairs on three scales: \texttt{goals, operations,} and \texttt{effects}. The judges had the option to choose whether they \texttt{agree, conditionally agree,} or \texttt{disagree} with confidence levels \texttt{high, medium}, and \texttt{low}.
   
    $\textbf{b)}$ \textbf{BigCloneBench dataset}. BigCloneBench (BCB) \cite{6976121} dataset, released by Svajlenko et al., was developed from IJAdataset-2.0\footnote{https://sites.google.com/site/asegsecold/projects/seclone}. IJAdataset contains $25K$ open-source Java projects and $365M$ lines of code. The authors have built the BCB dataset from IJaDataset by mining frequently used functionalities, such as bubble sort. The initial release of a BCB covers ten functionalities, including  $6M$ clone pairs and $260K$ non-clone pairs. The current release of the BCB dataset has about $8M$ tagged clones pair covering $43$ functionalities. Some recent code clone detection tools TBCCD \cite{10.1109/ICPC.2019.00021}, CDLH \cite{10.5555/3172077.3172312} has used the initial version of the BCB covering ten functionalities for their experiments.
    Hence, to present a fair comparison with TBCCD, we have also used the same version. 
    \par BCB dataset has categorized clone types into five categories: Type-1, Type-2, Strongly Type-3, Moderately Type-3, and Weakly Type-3+4 (Type-4) clones. Since there was no consensus on minimum similarity for Type-3 clones and it was difficult to separate Type-3 and Type-4 clones, the BCB creators categorized Type-3 and Type-4 clones based on their syntactic similarity. Thus, Strongly Type-3 clones have at least $70\%$ similarity at the statement level. These clone pairs are very similar and contain some statement-level differences. The clone pairs in the Moderately Type-3 category share at least half of their syntax but contain a significant amount of statement-level differences. The Weakly Type-3+4 code clone category contains pairs that share less than $50\%$ of their syntax. Tables  \ref{tab:statistics} and \ref{tab:Percentage of clone-types in BigCloneBench} summarises the data distribution of the BCB dataset.


\subsection{Experimental Procedure and Analysis}

\subsubsection{RQ1: How effective is \tool as compared to other state-of-the-art approaches?}
To answer this RQ, we compared two variants of \tool with TBCCD \cite{10.1109/ICPC.2019.00021}, a state-of-the-art clone detector that uses AST and tree-based convolutions to measure code similarity. We followed similar experimental settings as used by Yu et al. in TBCCD \cite{10.1109/ICPC.2019.00021}. 
To address this RQ, we used datasets from GCJ and BCB. We reserved $30\%$ of the dataset for testing, and the rest we used for training and validation. For the BCB dataset, we use the same code fragments from the related work  \cite{10.1109/ICPC.2019.00021, 10.5555/3172077.3172312}. We had used around $700K$ code pairs for training. For validation and testing, we used $300K$ code clone pairs each.
For the GCJ dataset, we had $440K$ clone pairs and $44M$ non-clone pairs. Due to the combinative nature of clones and non-clones, non-clone pairs rapidly outnumber the clone pairs. To deal with this imbalance in clone classes, we did downsampling for non-clone pairs using a reservoir sampling approach. This gives us $500K$ non-clone pairs and $440K$ clone pairs. We evaluated the following variants of \tool against TBCCD:
 \\\\
 $\textbf{1)}$ \textbf{Edge-Unified \tool (EU-\tool):} In this variant, we did not differentiate between the control and data-dependent edges to learn the program features. 
\\\\
$\textbf{2)}$ \textbf{Edge-Attributed \tool (EA-\tool)}: Program dependence graphs model control and data flow explicitly. Hence, it is logical to leverage this information as well while learning node representations. To model edge attributes, we have learned different program representations for data-dependent edges ($G_{data}$) and control-dependent edges ($G_{control}$) and aggregated them to obtain the final graph representation ($G_{final}$).
\subsubsection{RQ2: How well \tool generalizes on unseen projects and data sets?}
\label{rq:rq2}
To evaluate the robustness and generalizability of EU-\tool and EA-\tool, we evaluated the proposed approaches on unseen projects. 
In particular, we took EU-\tool, EA-\tool, and TBCCD trained on the GCJ dataset. We then tested the stability of the above tools on the following datasets:
\\\\
$\textbf{1)}$ \textbf{GoogleCodeJam (GCJ$^{*}$):}
We used the dataset of functionally similar code snippets (FSCs) proposed by Wagner et al. \cite{Wagner2016HowAF}. This dataset comprises of $32$ clone pairs from $\mathit{GCJ 2014}$. The authors classified the pairs into full syntactic similarity and partial syntactic similarity. The clone pairs are further classified into five categories $-$ \textit{Algorithms, Data Structures, Input/Output, Libraries, and Object-Oriented Design}. Each category has three different clone pairs classified based on the degree of similarity $-$ \textit{low, medium, and high}.
\\\\
$\textbf{2)}$ \textbf{SeSaMe Dataset:}
    This dataset  \cite{10.1109/MSR.2019.00079} has reported $857$ semantically similar clone pairs from $11$ open-source Java projects. However, of the $11$ projects, we were able to compile only eight projects, which gave us $93$ clone pairs for evaluation.

\section{Results}
\label{sec:results}
\begin{table}[t]
    \caption{Comparative evaluation with TBCCD variants.}
  \label{tab:Comparative evaluation with other state of the art approaches}\centering
\def\arraystretch{1.2}
  \begin{tabular}{|M{0.12\textwidth}|M{0.06\textwidth}
  |M{0.025\textwidth}|M{0.025\textwidth}|M{0.025\textwidth}
  |M{0.025\textwidth}|M{0.025\textwidth}|M{0.025\textwidth}|}
  \hline
    \multicolumn{1}{|c|}{\multirow{1}{*}{\textbf{Tool}}}&
     \multicolumn{1}{c|}{\multirow{1}{*}{\textbf{$\#$Params}}}&
    \multicolumn{3}{c|}{\multirow{1}{*}{\textbf{BCB}}}&
    \multicolumn{3}{c|}{\multirow{1}{*}{\textbf{GCJ}}}\\
    \cline{3-5} \cline{6-8}
   && \multicolumn{1}{c|}{\textbf{P}} & \multicolumn{1}{c|}{\textbf{R}} & 
   \multicolumn{1}{c|}{\textbf{F1}}  & \multicolumn{1}{c|}{\textbf{P}} &
    \multicolumn{1}{c|}{\textbf{R}} &
   \multicolumn{1}{c|}{\textbf{F1}} \\
   \hline
    TBCCD(-token) & $2.1\times10^{5}$ & $0.77$ & $0.73$ & $0.74$ & $0.77$ & $0.80$ & $0.80$\\
    \hline
    TBCCD & $1.7\times10^{5}$  & $0.96$ & $.96$ & $0.96$ & $0.79$ & $0.85$ & $0.82$\\
    \hline
    EU-\tool & $1.7\times10^{6}$ & $0.72$ & $0.97$ & $0.83$ & $0.84$ & $0.92$ & $0.88$\\ 
    \hline
    \textbf{ EA-\tool } & $6.6\times10^{6}$ & $\textbf{0.97}$ & $\textbf{0.98}$ & $\textbf{0.98}$ & $\textbf{0.91}$ & $\textbf{0.93}$ & $\textbf{0.92}$\\
    \hline
  \end{tabular}
    \end{table}
\begin{table}[t]
\def\arraystretch{1.2}
\centering \caption{F1 value comparison w.r.t various clones types in BigCloneBench dataset.}
\label{tab: F1 value comparison w.r.t various clones types in BigCloneBench dataset}\small
\begin{tabular}{|c|c|c|c|c|c|}
\hline
\textbf{Tools} & \textbf{T1} & \textbf{T2} & \textbf{ST3} & \textbf{MT3} & \textbf{WT3/T4}\\
\hline
TBCCD(-token) & $\textbf{1.0}$ & $0.90$ & $0.80 $& $0.65$ & $0.60$\\
\hline
TBCCD & $\textbf{1.0}$ &$\textbf{ 1.0}$ & $0.98$ & $0.96$ & $0.96$\\
\hline
EU-HOLMES & $\textbf{1.0}$ & $\textbf{1.0}$ & $0.86$ & $0.80$ & $0.80$\\
\hline
EA-HOLMES & $\textbf{1.0}$ & $\textbf{1.0}$ & $\textbf{0.99}$ & $\textbf{0.99}$ & $\textbf{0.99}$\\
\hline
\end{tabular}
\end{table}

\begin{lstlisting}[float=t,caption= A WT3/T4 clone example from BCB dataset. The code snippets are implementing the functionality for copying the directory and its content. Although the snippets are reported under Wt3/T4 clone category they are syntactically similar with some differences in the sequence of invoked methods and API calls. ,captionpos=b,label=t4clone,firstnumber=1,frame=tlrb]
    public void copyDirectory(File srcDir, File dstDir){
        if (srcDir.isDirectory()){
            if (!dstDir.exists())
                dstDir.mkdir();
            String[] children = srcDir.list();
            for (int i = 0; i < children.length; i++) {
                copyDirectory(new File(srcDir, children[i]), 
                new File(dstDir, children[i]));
            }
        }
        else{
            copyFile(srcDir, dstDir);
        }
    }
    //clone pair
    public static void copy(File src, File dst){
        if (src.isDirectory()) {
            String[] srcChildren = src.list();
            for (int i = 0; i < srcChildren.length; ++i) {
                File srcChild = new File(src, srcChildren[i]);
                File dstChild = new File(dst, srcChildren[i]);
                copy(srcChild, dstChild);
            }
        }
        else 
            transferData(src, dst);
    }
\end{lstlisting}
\begin{table}[t]
\def\arraystretch{1.2}
\centering
\caption{Performance on unseen dataset.}
\label{tab:Performance on unseen dataset}\centering\small

\begin{tabular}{|c|c|c|c|c|c|c|}
\hline
\multicolumn{1}{|c|}{\multirow{1}{*}{\textbf{Tool}}}&
\multicolumn{3}{c|}{\multirow{1}{*}{\textbf{GCJ$^{*}$}}}&
\multicolumn{3}{c|}{\multirow{1}{*}{\textbf{SeSaMe}}}
\\
\cline{2-4} \cline{5-7}
&\multicolumn{1}{c|}{\textbf{P}}&
\multicolumn{1}{c|}{\textbf{R}}&
\multicolumn{1}{c|}{\textbf{F1}}&
\multicolumn{1}{c|}{\textbf{P}} &
\multicolumn{1}{c|}{\textbf{R}}& 
\multicolumn{1}{c|}{\textbf{F1}} \\
\hline
TBCCD & $1.0$  &$0.63 $ & $0.77$ & $1.0$ & $0.48$  & $0.64$ \\
\hline
EU-\tool & $1.0 $& $0.65$  & $0.78$ & $1.0$ & $0.52$ & $0.68$ \\
\hline
\textbf{  EA-\tool }  & \textbf{$1.0$} & \textbf{$0.87$} & \textbf{$0.93$} & \textbf{$1.0$} & \textbf{$0.85$} & \textbf{$0.92$}\\
\hline
\end{tabular}%
\end{table}
\begin{figure*}[!t]
\begin{tabular}{p{0.5\textwidth}p{0.5\textwidth}}
\begin{minipage}{0.45\textwidth}
\centering
 \includegraphics[width=\textwidth]{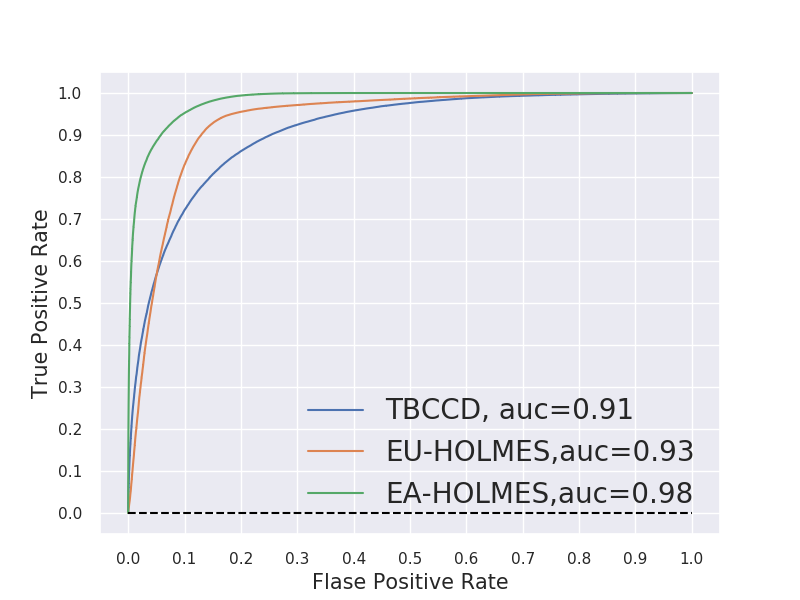}
        \caption{ROC curve of TBCCD, EU-\tool, EA-\tool on GCJ dataset. (AUC values are rounded up to $2$ decimal places)}
        \label{gcjval}
\end{minipage}
&
\begin{minipage}{0.45\textwidth}
\centering
 \includegraphics[width=\textwidth]{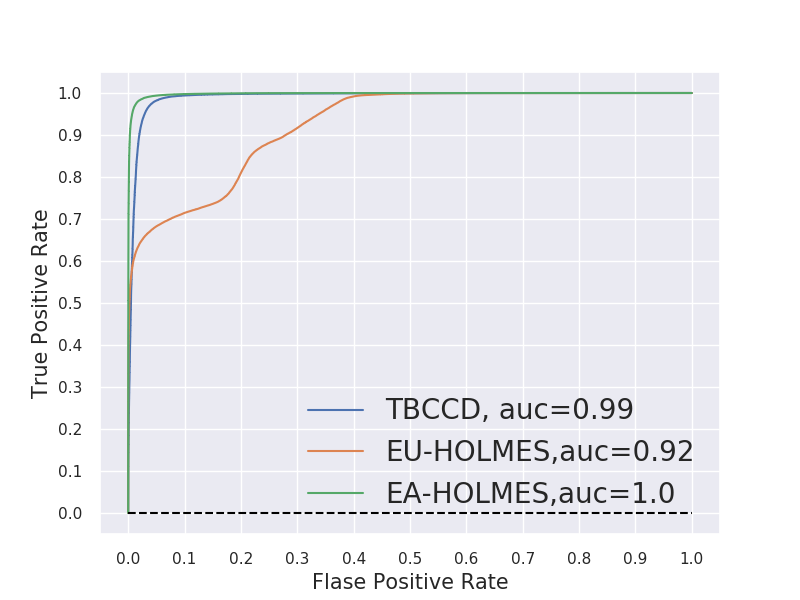}
        \caption{ROC curve of TBCCD, EU-\tool, EA-\tool on BCB dataset.(AUC values are rounded up to $2$ decimal places)}
        \label{bcbtest}
\end{minipage}
\end{tabular}
\end{figure*}
The results of our comprehensive evaluation are summarized in this section.

\begin{table*}[!tbp]
\def\arraystretch{1.2}
\caption{Detailed analysis of results on unseen GCJ$^{*}$ submissions.}
  \label{tab:Detailed analysis of results on unseen google code jam submissions}\centering\small
  \begin{tabular}{|c|c|c|c|c|c|c||c|c|c|c|c|c|c|}
\hline
    \multicolumn{1}{|c|}{\multirow{1}{*}{\textbf{Category}}}&
    \multicolumn{6}{c||}{\multirow{1}{*}{\textbf{  EA-\tool }}}&
    \multicolumn{6}{c|}{\multirow{1}{*}{\textbf{TBCCD}}}
    \\
    \cline{2-7} \cline{8-13}
    &\multicolumn{2}{c|}{\multirow{1}{*}{\textbf{Low}}}&
    \multicolumn{2}{c|}{\multirow{1}{*}{\textbf{Medium}}}&
    \multicolumn{2}{c||}{\multirow{1}{*}{\textbf{High}}}&
    \multicolumn{2}{c|}{\multirow{1}{*}{\textbf{Low}}}&
    \multicolumn{2}{c|}{\multirow{1}{*}{\textbf{Medium}}}&
    \multicolumn{2}{c|}{\multirow{1}{*}{\textbf{High}}}
   \\
   \cline{2-3} \cline{4-5} \cline{6-7} \cline{8-9}  
   \cline{10-11} \cline{12-13} 
    &\multicolumn{1}{c|}{\textbf{Full}}&
    \multicolumn{1}{c|}{\textbf{Part}}&
    \multicolumn{1}{c|}{\textbf{Full}}&
    \multicolumn{1}{c|}{\textbf{Part}}&
    \multicolumn{1}{c|}{\textbf{Full}}&
    \multicolumn{1}{c||}{\textbf{Part}}&
    \multicolumn{1}{c|}{\textbf{Full}}&
    \multicolumn{1}{c|}{\textbf{Part}}&
    \multicolumn{1}{c|}{\textbf{Full}}&
    \multicolumn{1}{c|}{\textbf{Part}}&
    \multicolumn{1}{c|}{\textbf{Full}}&
    \multicolumn{1}{c|}{\textbf{Part}}
   \\
    \cline{2-2}
    \cline{3-3} \cline{4-4} \cline{5-5} \cline{6-6} \cline{7-7} 
    \cline{8-8} \cline{9-9}     \cline{10-10} \cline{11-11}     \cline{12-12}  \cline{13-13}
    Data Structure & \checkmark & \checkmark & \checkmark & \ding{53} & \ding{53} & \checkmark &
    \ding{53} & \ding{53} & \checkmark & \checkmark & \ding{53} & \ding{53}
    \\
    \hline
    OO Design & \checkmark & \ding{53} & \checkmark& \checkmark & \checkmark & \checkmark &
    \checkmark & \ding{53} & \checkmark & \checkmark& \checkmark & \checkmark
    \\
    \hline
    Algorithm & \checkmark & \checkmark & \checkmark& \checkmark & \checkmark & \checkmark & \ding{53} & \checkmark & \checkmark & \ding{53} & \ding{53} & \checkmark\\
    \hline
    Library & \checkmark & \checkmark & \checkmark& \checkmark & \checkmark & \checkmark & \ding{53} & \checkmark& \checkmark& \checkmark& \ding{53}& \ding{53}\\
    \hline
    Input/Output & \checkmark & \checkmark & \ding{53}& \checkmark & \checkmark &  \checkmark &  \ding{53} & \checkmark & \checkmark  & \checkmark & \checkmark & \ding{53}\\
    \hline
  \end{tabular}
\end{table*}

\subsection{RQ1: How effective is \tool as compared to other state-of-the-art approaches?}
To answer this RQ, we compared our proposed approach variants EU-\tool and EA-\tool with two variants of TBCCD - $(1)$ TBCCD(-token), and $(2)$ TBCCD. These variants of TBCCD are reported in the paper \cite{10.1109/ICPC.2019.00021}. The variant TBCCD(-token) uses randomly initialized AST node embeddings in place of source code tokens, which are fine-tuned during training. The second variant, TBCCD, uses the token-enhanced AST and PACE embedding technique. The token-enhanced AST contains source code tokens such as constants, identifiers, strings, special symbols, etc. Table \ref{tab:Comparative evaluation with other state of the art approaches} shows the comparative evaluation of EU-\tool and EA-\tool with TBCCD(-token) and TBCCD on the BCB and GCJ datasets.
\par On the BCB dataset from Table \ref{tab:Comparative evaluation with other state of the art approaches}, we can see both TBCCD and EA-\tool perform equally well, while the performance of TBCCD(-token) drops significantly. The BCB dataset categorizes clones into five categories: Type-1 clones, Type-2 clones, Strongly Type-3 clones, Moderately Type-3 clones, and Weakly Type-3+4 (Type-4) clones. Since there was no consensus on minimum similarity for Type-3 clones, and it was difficult to separate Type-3 and Type-4 clones, the BCB creators categorized Type-3 and Type-4 clones based on their syntactic similarity. Thus, Strongly Type-3 clones have at least $70\%$ similarity at the statement level. These clone pairs are very similar and contain some statement-level differences. The clone pairs in the Moderately Type-3 category share at least half of their syntax but contain a significant amount of statement-level differences. The Weakly Type-3+4 code clone category contains pairs that share less than $50\%$ of their syntax. Table \ref{tab: F1 value comparison w.r.t various clones types in BigCloneBench dataset} further shows the performance of TBCCD(-token), TBCCD, EA-\tool, and EU-\tool on different code clone types in the BCB dataset. All the approaches achieve good performance on Type-1 and Type-
2 code clone categories, as these code clone types are easier to detect. While on the hard-to-detect code clone categories such as Moderately Type-3, Weakly Type-3+4, TBCCD(-token) performs poorly compared to the TBCCD variant, we also see an improvement of {\raise.17ex\hbox{$\scriptstyle\mathtt{\sim}$}}$3\%$ in EA-\tool as compared to TBCCD. The reason for the improved performance of TBCCD is attributed to the use of syntactic similarity existing between the code snippets in the BCB dataset, as shown in Listing \ref{t4clone}. This syntactic similarity existing in the form of identifiers, tokens, etc., is exploited by TBCCD while learning for code similarity. 
\par On the other hand, on the GCJ dataset, the performance of both TBCCD's variants, i.e., TBCCD(-token) and TBCCD, drops significantly. We can see an improvement of {\raise.17ex\hbox{$\scriptstyle\mathtt{\sim}$}}$10\%$ in the F1-score on the GCJ dataset in EA-\tool compared to TBCCD. The performance drop in TBCCD(-token) and TBCCD on the GCJ dataset is attributable to two factors: $(1)$ Both TBCCD's variants use ASTs to learn program features. ASTs represent program syntactic structure, and as shown in Listings \ref{code1} and \ref{code2}, the code clone pairs in GCJ have a significant structural difference. Thus, the ASTs of these code pairs are very different, making it hard for the model to infer similarity; and  $(2)$ As opposed to the BCB dataset, where there was some syntactic similarity between the code pairs, the GCJ code clone pairs have substantial differences in structure and algorithm. These differences are not unexpected because the submissions are made by independent programmers implementing the solutions from scratch. Consequently, without modeling semantics, the GCJ dataset's clones are harder to detect compared to BCB.
\par Also, from Tables \ref{tab:Comparative evaluation with other state of the art approaches} and \ref{tab: F1 value comparison w.r.t various clones types in BigCloneBench dataset}, we observe that EA-\tool performs better than EU-\tool on GCJ and BCB datasets in every evaluation metric. This performance difference demonstrates the importance of structured semantics of source code while learning code functional similarity.
\par Additionally, to analyze the diagnostic ability of TBCCD, EU-\tool, and EA-\tool, we plotted the Receiver Operating Characteristics (ROC) curve by varying the classification threshold. Figures \ref{gcjval} and \ref{bcbtest} show the ROC curve and corresponding Area Under Curve (AUC) values for the GCJ and BCB dataset. We have plotted the ROC curve of the TBCCD variant only, as it has outperformed the TBCCD(-token) variant on the GCJ and BCB datasets. For all other experiments also, we have considered the TBCCD variant only. 
\par ROC curve is a graphical plot, visualizing trade-off between True Positive Rate (TPR) plotted on the y-axis and False Positive Rate (FPR) plotted on the x-axis. AUC metric measures the degree of separability. Generally, an excellent classifier has a high AUC, denoting the model is better at predicting clone pairs as clones and non-clone pairs as non-clones. We can see from Figures \ref{gcjval} and \ref{bcbtest} that EA-\tool has the best AUC value on both the datasets.
\begin{figure}[t]
    \centering
 \includegraphics[width=0.5\textwidth]{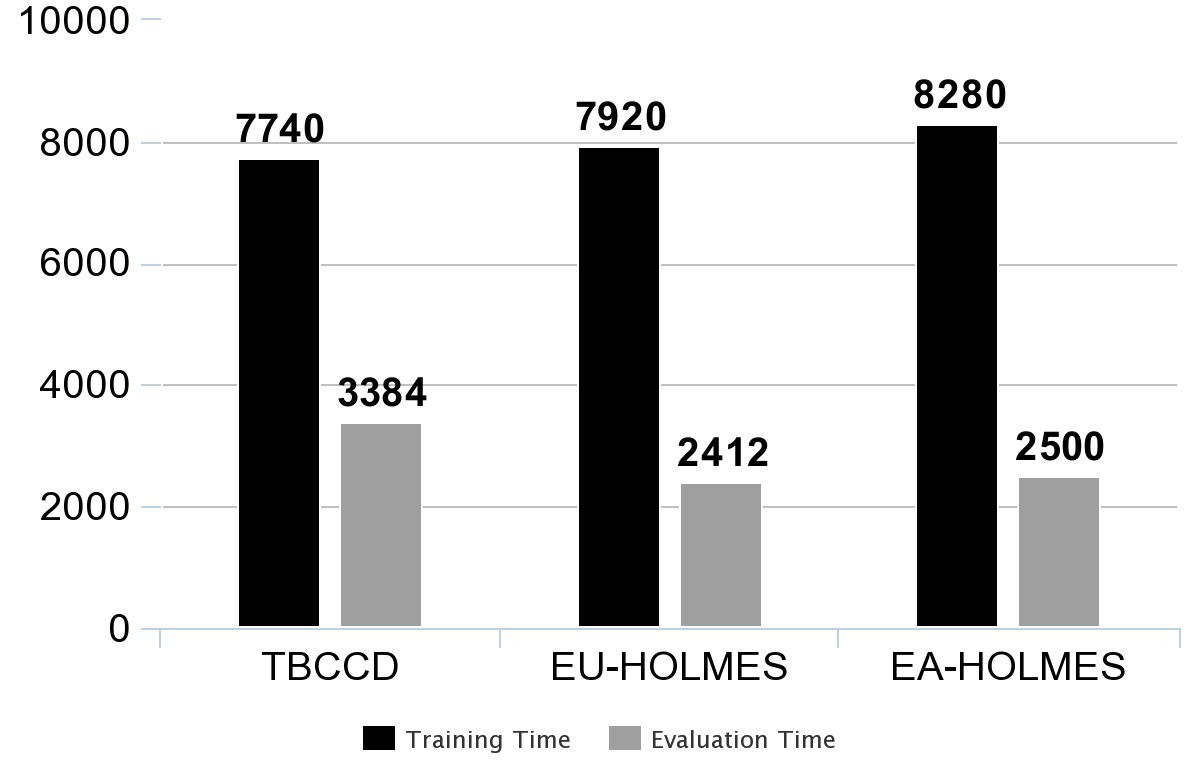}
    \caption{Time performance analysis on the GCJ dataset.}
    \label{Time performance analysis on GCJ dataset}
\end{figure}
\par \textbf{Time Performance}. We also evaluated the time performance of TBCCD, EU-\tool, and EA-\tool on two parameters $- (1)$ time taken to build ASTs vs. time taken to build PDGs, and $(2)$ total training and evaluation time. We run each of these tools with the same parameter settings reported in Section \ref{subsec:parmeter} on the full GCJ dataset on a Workstation with an Intel Xeon(R) processor having $24$ CPU cores. We have used a GeForce RTX $2080$Ti GPU with $11 GB$ of GPU memory.
\par The total time taken to build AST for $9436$ project files was $30$ minutes, while PDG took $60$ minutes. Figure \ref{Time performance analysis on GCJ dataset} shows the training and evaluation time analysis of TBCCD, EU-\tool, and EA-\tool. EA-\tool learns separate representation for the control and data dependence graphs. Thus, it takes more training time than EU-\tool and TBCCD. Even though the total time taken to build PDGs is greater than building ASTs and EA-\tool takes longer training time, these are one-time offline processes. Once a model is trained, it can be reused to detect code clones.
\subsection{RQ2: How well \tool generalizes on unseen projects and data sets?}
Table \ref{tab:Performance on unseen dataset} shows performance of EU-\tool, EA-\tool, and TBCCD on unseen datasets. We can see that EA-\tool performs considerably better on both datasets as compared to TBCCD.

\par Table \ref{tab:Detailed analysis of results on unseen google code jam submissions} shows the detailed classification result of TBCCD and EA-\tool on the GCJ$^{*}$ dataset. In the table, $\checkmark$ indicates that the code clone detector detects the pair, while \ding{53} indicates that the pair goes undetected. From table \ref{tab:Detailed analysis of results on unseen google code jam submissions}, we can infer that our proposed approach can detect the majority of the pairs correctly, even the pairs with partial syntactic similarity. These results affirm that EA-\tool generalizes well on unseen datasets also.


\section{Discussion}
\label{sec:discussion}
\begin{figure}[t]
    \centering
  \includegraphics[width = 7.5cm, height=5.6cm]{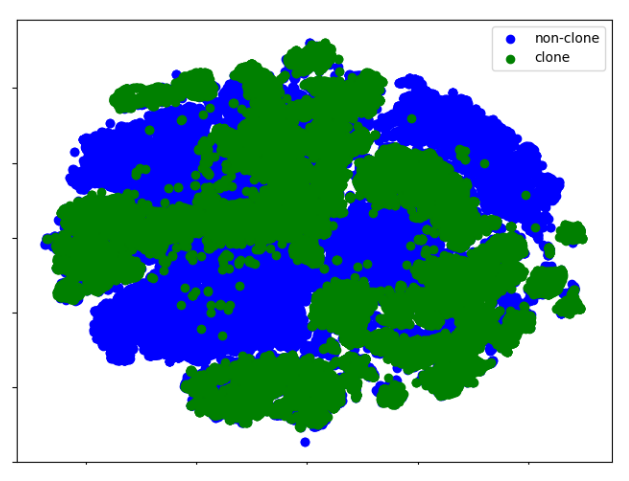}
        \caption{t-SNE plot of graph embeddings of clone and non-clone pairs of GCJ dataset generated by EA-\tool.}
        \label{final hidden layer embedding}
\end{figure}
\begin{figure*}
\begin{tabular}{p{0.4\textwidth}p{0.5\textwidth}}
\begin{subfigure}{0.4\textwidth}
 \includegraphics[width=\textwidth,height=6cm]{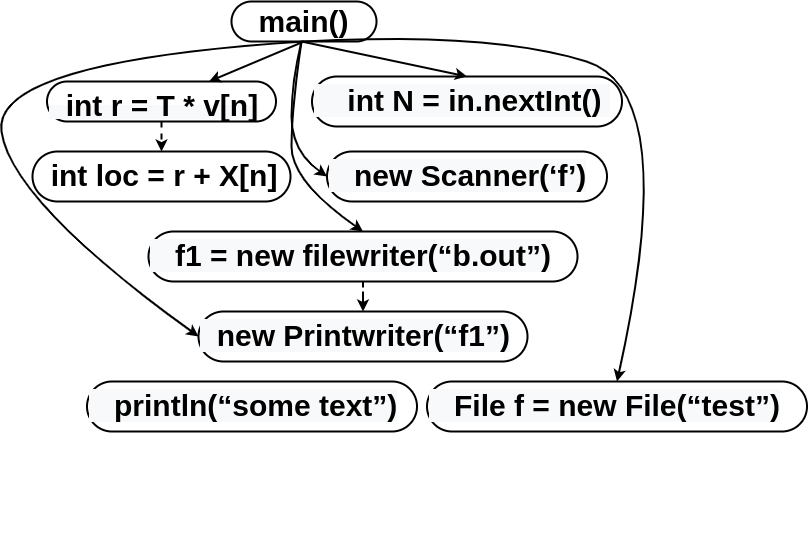}
        \caption{PDG for a java source code. Solid line edges denote control dependence. Dashed line edges denote data dependence.}
        \label{fig:pdgfeatures}
\end{subfigure}
&
\begin{subfigure}{0.5\textwidth}
 \includegraphics[width=\textwidth]{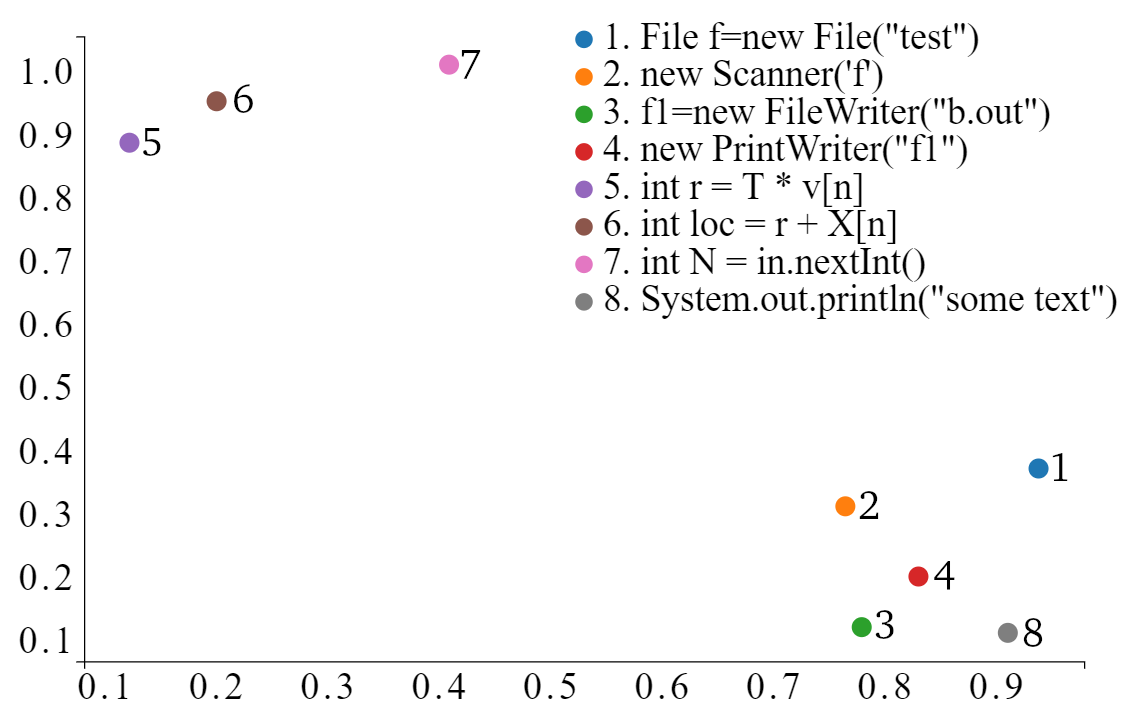}
        \caption{t-SNE plot of input feature vectors generated by EA-\tool.}
        \label{fig:input feature vectors}
\end{subfigure}

\end{tabular}
\caption{Qualitative analysis of the features learned by EA-\tool.}
\end{figure*}
\begin{figure*}[h]
\begin{subfigure}{.4\textwidth}
\begin{lstlisting}[firstnumber=1]
    package test;
    import java.util.*;
    public class Test{  
    public static void main(String[] args)
    {
        Scanner input = new Scanner(System.in);
        int T = input.nextInt(); 
        int[] a = new int[T];
        for (int j = 0; j < T; j ++) 
            a[j] = input.nextInt();
        int c = 0;
        for (int j = 0; j < T; j ++) 
            if (a[j] == j+1) 
                c ++;
    }
}
\end{lstlisting}
\caption{Java Source Code.}
\label{javasourcecode}
\end{subfigure}
  \begin{subfigure}{.6
  \textwidth}
\centering
  \includegraphics[height=5.6cm,width=8cm]{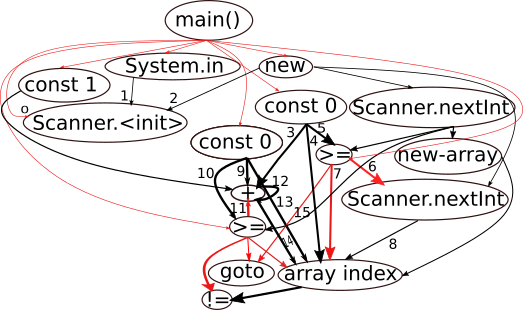}
      \caption{Attention Encoded Program Dependence Graph. Control dependence edges are colored red, whereas data dependent edges are colored black.}
\label{attention encoded pdg}
\end{subfigure}
\caption{Qualitative assessment of the learned PDG paths. }
\end{figure*}

\begin{figure*}
    \centering
    \includegraphics[width=\textwidth]{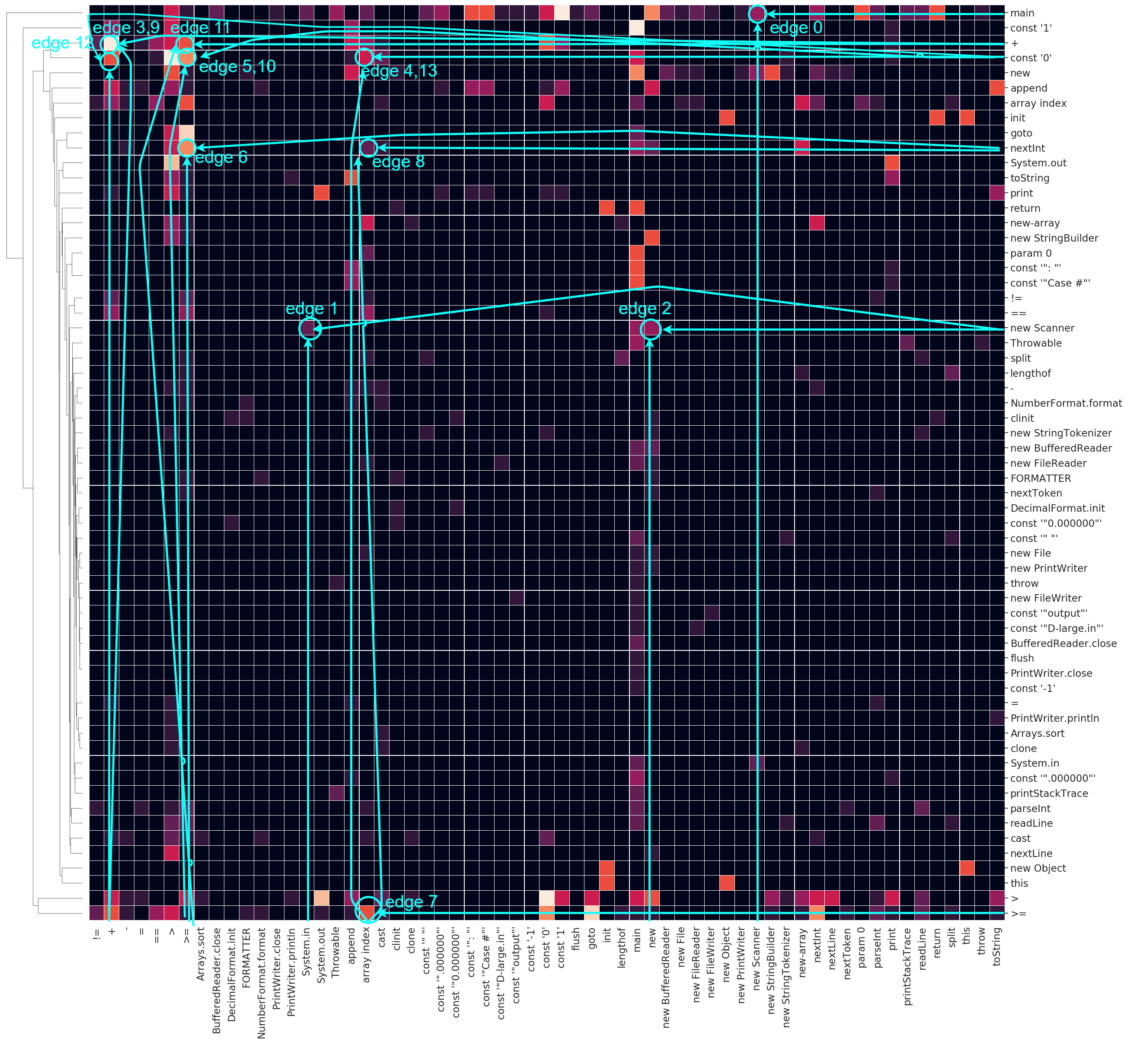}
    \caption{Cluster map of the attention scores received by code snippets similar to Listing \ref{javasourcecode}. The annotations in Figure denote the corresponding edges of Figure \ref{attention encoded pdg}. [Best viewed in color.]}
    \label{fig:attentionclustermap}
\end{figure*}
\begin{figure}
    \centering
    \includegraphics[width=0.5\textwidth]{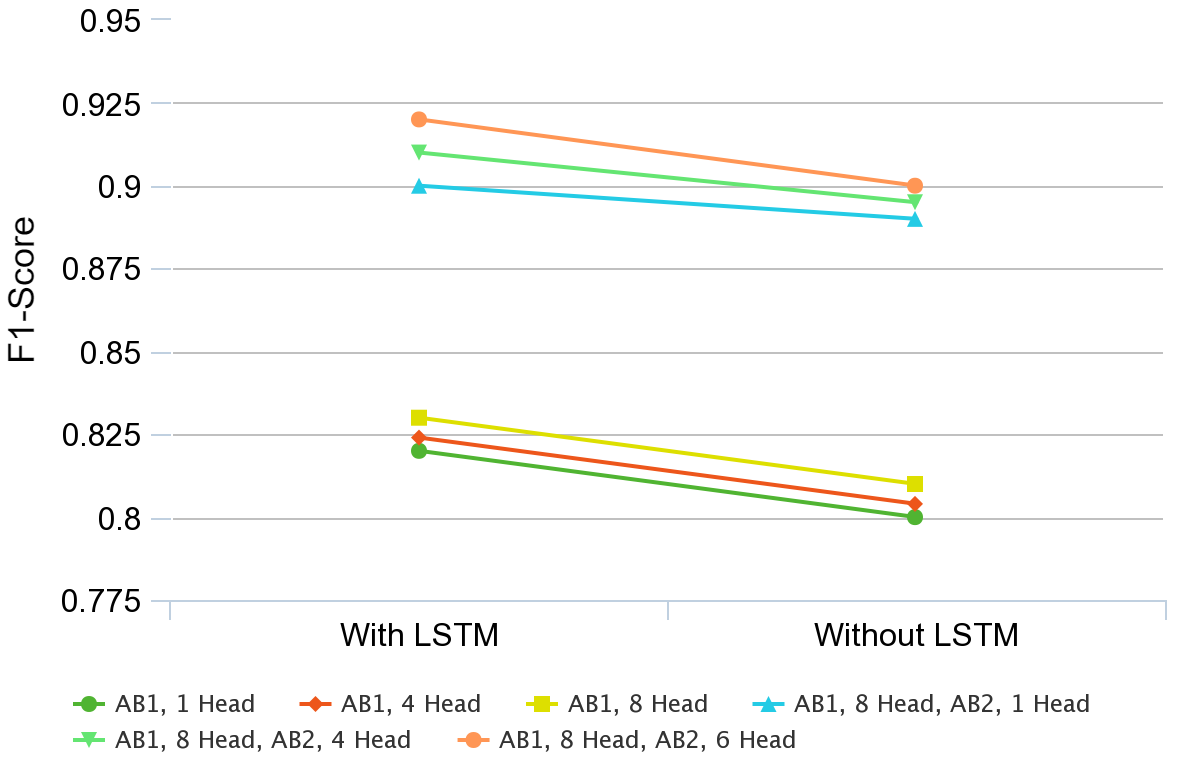}
    \caption{The effect on the performance of \tool after varying the number of attention heads in both the attention blocks and after removing the LSTM layer. AB in the legend stands for Attention Block, for instance, “AB1, 1 Head” corresponds to “Attention Block 1 and 1 attention head”.}
    \label{fig:attnlstmexpt}
\end{figure}
\begin{figure}
    \centering
    \includegraphics[width=0.5\textwidth]{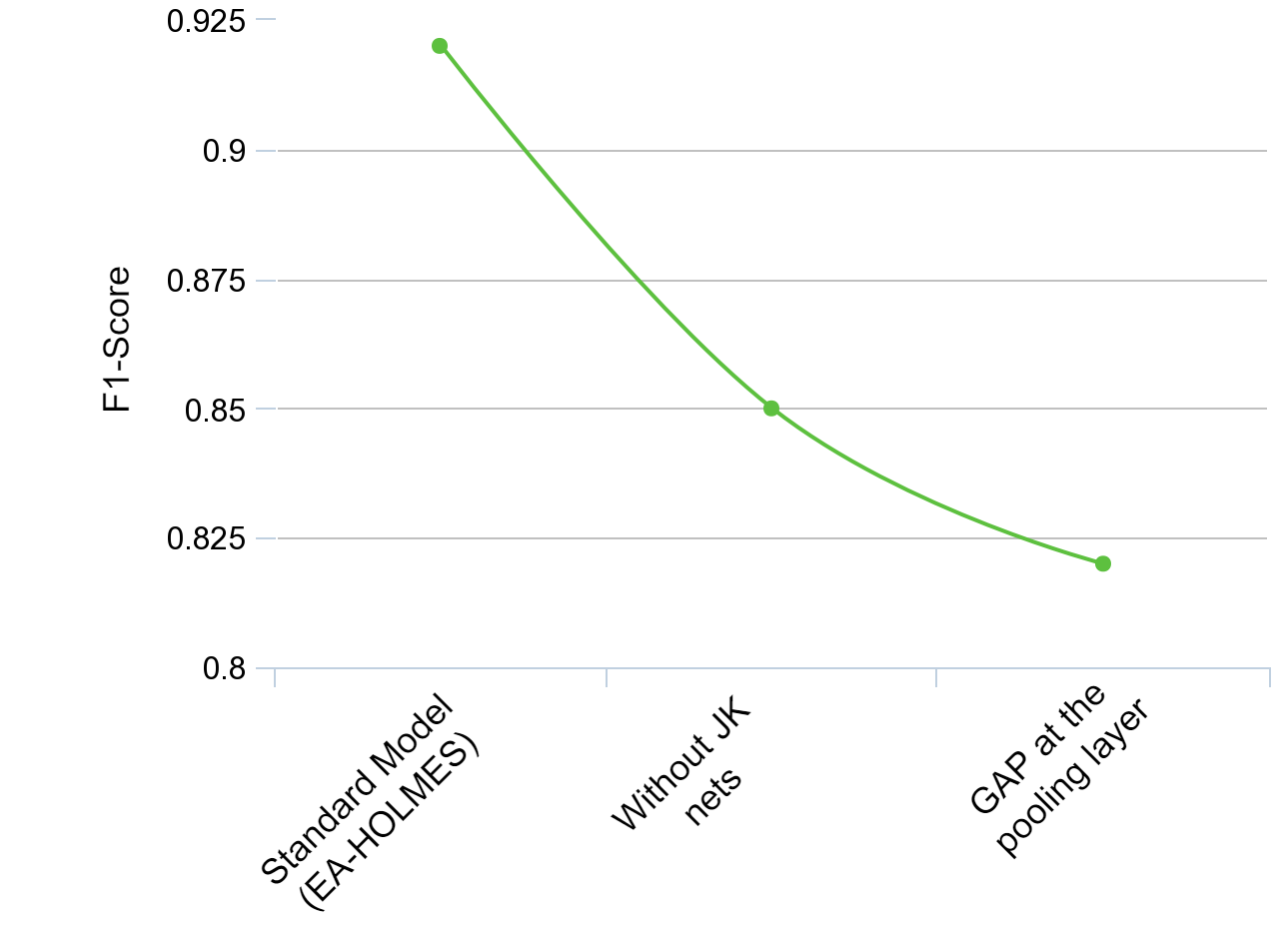}
    \caption{The effect on the performance of \tool after removing Jumping Knowledge (JK) nets and soft attention mechanism from the graph readout layer.}
    \label{fig:jkgapexpt}
\end{figure}

\subsection{Why \tool outperforms other state-of-the-art clone detectors}
Our approach uses Program Dependence Graphs (PDGs) for representation learning. PDGs represent the program's semantics through data dependence and control dependence edges. Our approach models the relations between the program statements in PDGs using Graph Neural Networks (GNNs). We have used attention-based GNNs and LSTMs to filter and aggregate relevant paths in PDGs that enable us to learn semantically meaningful program representations. Attention-based GNNs draw importance to different direct (one-hop) neighbors, while LSTMs are used to capture wider context and long-range dependencies between nodes of the PDG.
\par Thus representing source code as graphs and modeling them through GNNs and LSTMs helps us to leverage the program's structured semantics, contrary to using ASTs and token sequences for learning program features. Additionally, to give respective importance to control and data dependence relations between different statements, we learned two different representations corresponding to each edge type. This information helps us to differentiate and prioritize between the available semantic and syntactic relations between different program statements.
\subsection{Representation learning using Graph Attention networks (GATs).}
Though there are many graph feature learning layers in the literature, such as Graph Convolutional Networks (GCN), Gated Graph Neural Network (GGNN), our work uses the GAT layer to learn program dependency graph nodes' features. This is because the GAT layer can learn an adaptive receptive path for a node in a graph, i.e., assigning different importance to different nodes in the same neighborhood. On the other hand, the GCN layer has fixed receptive paths, which might not work well in our case as all paths in the PDG are not equally important. The importance of attention has also been demonstrated in Figures \ref{attention encoded pdg} and \ref{fig:attentionclustermap}. Using GGNN, another recurrent graph feature learning layer, can also be problematic as it uses Gated Recurrent Unit (GRU) and Backpropagation Through Time (BPTT),  which can be problematic for large graphs and may require large memory.
\subsection{Qualitative analysis of the features learned by \tool}
\label{subsec:qualitativeanalysisofthefeatureslearnedbyholmes}
Figure \ref{final hidden layer embedding} shows the t-SNE \cite{Maaten08visualizingdata} plot of the final hidden layer of \tool trained on the GCJ dataset. The figure shows that the learned features can effectively differentiate between the clone and non-clone classes. Since we have only achieved the F1-Score of $92\%$ on GCJ dataset, we can see some overlap between the clone and non-clone classes in the figure.
\par To get more insights into the learned feature space of \tool, we processed and extracted the node features and adjacency matrix from the PDG shown in Figure \ref{fig:pdgfeatures}. We then passed this to the first hidden layer of \tool. Figure \ref{fig:input feature vectors} shows the t-SNE visualization of the generated vector embedding.
\par It can be seen from the figure that the statements that share similar semantics are plotted very closely. In contrast, the statements that are not similar are not close in the embedding space. For instance, the statements $int \; r = T * v[n]$ and $int\; loc  = r  +x[n]$ are plotted nearby, as the latter uses the former's result, and both are performing some numerical computation. Similarly, statements $1$,$2$,$3$,$4$ and $8$ are similar, thus plotted nearby in embedding space. These insights suggest that our approach models graph topology and node distribution simultaneously for learning the graph representation.
\subsection{Qualitative assessment of the learned PDG paths}
To gain further insights into our model's working, we plotted the aggregate of multi-head self-attention scores on the PDG paths. Figure \ref{attention encoded pdg} shows the plot of attention scores received by the PDG of Listing \ref{javasourcecode}. Here in Figure \ref{attention encoded pdg}, the edge thickness denotes the attention score received by each path. From the Figure, it can be seen that the model assigns higher weights to the semantic paths. 
\par For instance, in Listing \ref{javasourcecode}, statement $6$ initializes the \texttt{Scanner} class's object from the \texttt{java.util} package. The Scanner class is used to read input in a java program. The attention scores received by statement $6$ are shown by edges $0-2$ in Figure \ref{attention encoded pdg}. In the same code, the attention scores received by statement $9-10$ and $12-14$ are shown through edges $3-8$ and $9-15$, respectively. Statement $9-10$ from Listing \ref{javasourcecode} initializes an array of size T with random integers, and statement $12-14$ increments variable c based on some condition. From Figure \ref{attention encoded pdg}, it can also be seen that statement $6$, being the general object creation statement of \texttt{java.util.Scanner} class, receives less attention as compared to statements $9-10$ and $12-14$. Even the statement 10 that fills the array with some random runtime integers received less attention, as shown through edge $8$. This suggests that \tool learns to attend important semantically meaningful paths that contain the actual application logic.
\par Further, to strengthen the above claims, we randomly selected five Java programs similar to Listing \ref{javasourcecode} and computed attention scores received by each PDGs of these programs. We then aggregated the attention scores received by these snippets across similar paths. Figure \ref{fig:attentionclustermap} shows the cluster map \cite{seaborn} of the attention scores received by these Java programs.
\par A cluster map is a clustered version of a heatmap that uses hierarchical clustering to order data by similarity. In Figure \ref{fig:attentionclustermap}, we have used cosine similarity across the rows to group similar statements together. The brighter color in the Figure represents higher attention and vice versa. From Figure \ref{fig:attentionclustermap}, we can see that the attention received across all five snippets is consistent with the attention scores received in Figure \ref{attention encoded pdg}. For instance, the attention scores received on edges $0-2$ (statement initializing \texttt{Scanner} class object) is less than the attention scores received on edges $3-8$ (array initialization statement) as shown in Figure \ref{attention encoded pdg} and this can also be verified in Figure \ref{fig:attentionclustermap} (the edges are annotated in the Figure). Similarly, the attention scores received on edges $9-13$ in Figure \ref{attention encoded pdg} are similar to the attention score received by the five random java snippets' attention score, as shown in Figure \ref{fig:attentionclustermap}.
\par Besides this, the cluster map in Figure \ref{fig:attentionclustermap} also justifies our claims made in Section \ref{subsec:qualitativeanalysisofthefeatureslearnedbyholmes}. As we can see, relational operators such as $>$, $>=$, !=, and == are clubbed together across rows of Figure \ref{fig:attentionclustermap}. File handling methods like the buffered reader, file reader, are also clubbed together. FileWriter is clubbed with PrintWriter and new File statements. Thus we can say that \tool learns to attend semantically meaningful paths along with modeling graph topology and node distributions.
\subsection{Ablation Study}
To understand the contribution of each component of our model, we conducted an ablation study, and the results are shown in Figures \ref{fig:attnlstmexpt} and \ref{fig:jkgapexpt}. In the first set of experiments, we varied the number of attention heads of both the attention blocks while keeping the rest of the architecture the same. The left part of Figure \ref{fig:attnlstmexpt} shows the results of varying attention heads. From Figure \ref{fig:attnlstmexpt}, we can see that the \tool performance degrades on removing the second attention block. The F1-score also degrades further when we reduce the number of attention heads. 
\par Next, to examine the influence of LSTMS on the model’s performance, we removed the LSTM layer from the \tool architecture and varied the attention heads of both the attention blocks. The results are shown in the right part of Figure \ref{fig:attnlstmexpt}. We can see that after removing the LSTM layer from the \tool architecture, the performance degrades. This shows the importance of using LSTMS in aggregating the local neighborhood for learning node representation.
\par We removed the Jumping Knowledge (JK) networks from the \tool architecture in the next set of experiments. The results in Figure \ref{fig:jkgapexpt} show that the F1-score reduces drastically after removing JK nets. Thus, it can be said that the JK nets help in the model’s stability and improve performance. In the end, we replaced the soft attention mechanism with the Global Average Pooling (GAP) layer to learn graph representation. GAP layer simply averages all the learned node representations to make up the final hidden graph representation. From Figure \ref{fig:jkgapexpt}, we can observe that \tool performance degrades when a GAP layer is employed in place of the soft attention mechanism. This shows the importance of the soft attention mechanism at the graph readout layer.
\subsection{Limitations of \tool}
We have used PDG representation to learn the program features. Static analysis is required to generate PDGs, and it only works for compilable code snippets. Therefore, we cannot directly apply our technique to incomplete programs. For this reason, we have used JCoffee \cite{Jcoffee} to infer the missing dependencies in the BCB dataset. In addition, we have used a supervised learning approach to learn code similarity, which is expensive in terms of labeled dataset procurement. However, as shown in the results, our model can learn a generalized representation and perform satisfactorily on unseen datasets. In our future work, we plan to extend our model with techniques such as domain adaptation and transfer learning so that it can be applied to other unseen and unlabeled datasets.



\section{Threats To Validity}
\label{threats to validity}
\subsection{Implementing baselines on our datasets.}
We have used the available implementation of TBCCD \cite{10.1109/ICPC.2019.00021}. 
There are various options available to tune the hyper-parameters of TBCCD, such as varying batch size, learning rate, etc. Each possible option tuning of TBCCD might have produced different results. To mitigate this, we have selected the default settings (the best parameters for TBCCD) reported by Yu et al. in \cite{10.1109/ICPC.2019.00021}.
\subsection{Generalizing results in other programming languages.}

In this paper, we have implemented the proposed approach for the Java language. While the PDG generation part is implemented in Java, all other subsequent steps are language agnostic. Attention, graph-based neural networks have been used in different contexts and for other languages as well. Also, the PDG can be generated for code written in other languages;  for instance, LLVMs can generate PDGs for C/C++ code snippets. Therefore, \tool can potentially be adapted to work for code written in other programming languages. However, since we have not tested this, we can not make a sound claim regarding its efficacy.
\subsection{Evaluating \tool on open source projects and programming competition datasets.}

We conducted experiments on two widely used datasets for code clone detection in this work - GoogleCodeJam and BigCloneBench. We have also tried to use a large and representative dataset for our experiments. Unlike the past work \cite{10.1145/3236024.3236068}, which has used $12$ different functionality in their evaluation, we have used $100$ different functionalities from GoogleCodeJam. However, \tool performance might vary across other projects, as these benchmarks are not representative of all software systems. To mitigate this threat and assess \tool generalizability, we have also performed some cross dataset experiments on SeSaMe and a GoogleCodeJam dataset variant.


\section{Related Work}
\label{sec:related work}
This section describes the related work on code clone detection techniques and learning program representations using learning-based techniques.
\subsection{Code Clone Detection}
   \subsubsection{Traditional code clone detection approaches.}
    \par Most traditional code clone detection techniques target Type 1-3 code clones. These techniques measures code similarity by using program representation such as abstract syntax trees \cite{deckard}, lexical tokens \cite{ccfinder,cpminer}, program text\cite{dup,nicad}. Deckard \cite{deckard}, a popular tree-based code clone detection technique, computes characteristic vector for AST nodes of the given program. It then applies Locality Sensitive Hashing (LSH) to find similar code pairs. SourcererCC \cite{7886988} is a token-based code clone detection technique that compares token subsequences to identify program similarity.
    \par There are also some graph-based techniques \cite{pdgdup,duplix} that use program dependence graphs to identify Type-4 clones. PDG-DUP \cite{pdgdup} first converts the given program to PDGs and then uses program slicing and subgraph isomorphism to identify clone pairs. DUPLIX \cite{duplix} also uses program slicing and graph isomorphism to identify similar code pairs. However, these approaches do not scale to large codebases and are very time-consuming, limiting their applications in practical software systems. In addition to these, some techniques also exist that compare program runtime behavior \cite{coderelatives} or program memory states \cite{6032469} to identify code clones.
\subsubsection{Learning based code clone detection approaches.}

Learning from data to identify code clones has been a great deal of interest from the past. There have been techniques using data mining approaches to learn code similarity \cite{Marcus, SK}. For example, Marcus and Maletic \cite{Marcus} has proposed to use latent semantic indexing to detect semantic clones. The proposed approach examines the source code text (comments and identifiers) and identifies the implementation of similar high-level concepts such as abstract data types. Much recent work uses learning-based techniques to learn code similarity. These approaches try to learn continuous vector-based representations of code fragments. These vectors are then compared using some distance metric (e.g., Euclidean distance) or using neural networks to measure code functional similarity.
White et al. \cite{10.1145/2970276.2970326} used a recursive neural network to learn program representation. They represented source code as a stream of identifier and literals and used it as an input to their deep learning model. Tufano et al. \cite{8595238} using a similar encoding approach as \cite{10.1145/2970276.2970326} encoded four different program representations- identifiers, Abstract Syntax Trees, Control Flow Graphs, and Bytecode. They then used a deep learning model to measure code similarity based on their multiple learned representations. Wei et al. \cite{ijcai2018-394} uses AST and tree-based LSTM to learn program representation. To incorporate structural information available with the source code Yu et al. \cite{10.1109/ICPC.2019.00021} uses tree-based convolutions over ASTs to learn program representation. Saini et al. \cite{Saini2018OreoDO} proposes using software metrics and machine learning to detect clones in the twilight zone. Zhao et al. \cite{10.1145/3236024.3236068} used feature vectors extracted from the data flow graph of a program to learn program representation using deep neural networks. Mathew et al. \cite{Mathew2020SLACCSL} proposed a cross-language clone detection by comparing the input and output of the potential clone candidates.Additionally, there also exists techniques to detect clones in binaries \cite{Gemini,GMN,DBLP:conf/iwpc/HuZLG17,DBLP:conf/icsm/HuZLWLG18}. Li et al. \cite{GMN} proposed a Graph Matching Network(GMN) to address the problem of matching and retrieval of graph structured data. They have proposed a new cross-graph attention-based matching mechanism to compute similarity score for a given pair of graph. The proposed graph matching network model is shown to outperform the graph embedding models on binary function similarity search. Xu et al. \cite{Gemini} proposed a technique to detect cross-platform clones in binaries. The proposed tool Gemini uses Structure2vec \cite{struct2vec} neural network model to learn the hidden binary function features from control flow graphs. The learned features are then compared using cosine distance to measure code similarity.
\subsection{Representation Learning for Source Code}
There has been a significant interest in utilizing deep learning models to learn program embeddings. The goal is to learn precise representations of source code for solving various software engineering tasks. Gupta et al. \cite{deepfix} propose to fix common programming errors using a multilayered sequence to sequence neural networks with attention. The deep learning model is trained to predict the erroneous program locations in a C program and the required correct statements. Allamanis et al. \cite{allamanis2017learning} use graph-based neural networks over AST based program representation to learn program embeddings. The learned embeddings are then used to predict the names of variables and varmisue bugs. Wang et al. \cite{wang2019learning} use program execution traces to learn program embeddings to predict method names. Ben-Nun et al. \cite{bennun2018neural} use Intermediate Representation (IR) of source code with recurrent neural networks to learn program embeddings. Hoang et al. \cite{hoang2020cc2vec} propose a neural network model CC2Vec to learn a representation of source code changes. CC2Vec uses attention to model the hierarchical structure of source code. The learned vectors represent the semantic intent of code change. The authors have evaluated the proposed approach on three downstream tasks: log message generation, bug fixing patch identification, and just-in-time defect prediction. There has also been some work on assessing the quality of learned representations. Kang et al. \cite{8952475} present an empirical study to assess the generalizability of Code2vec token embeddings. The authors have evaluated the Code2vec token embeddings on three downstream tasks: code comments generation, code authorship identification, and code clone detection. Their results show that the learned representation by the Code2vec model is not generalized and cannot be used readily for the downstream tasks.

\section{Conclusion and Future Work}
\label{sec:conclusion}
There has been a significant interest in detecting duplicated code fragments due to its pertinent role in software maintenance and evolution. Multitudinous approaches have been proposed to detect code clones. However, only a few of them can detect semantic clones. The proposed approaches use syntactic and lexical features to measure code functional similarity. They do not fully capitalize on the available structured semantics of the source code to measure code similarity. In this paper, we have proposed a new tool \tool for detecting semantic clones by leveraging the semantic and syntactic information available with the program dependence graphs (PDGs). Our approach uses a graph-based neural network to learn program structure and semantics. We have proposed to learn different representations corresponding to each edge-type in PDGs. 
\par We have evaluated both variants of \tool on two large datasets of functionally similar code snippets and with recent state-of-the-art clone detection tool TBCCD \cite{10.1109/ICPC.2019.00021}. Our comprehensive evaluation shows that \tool can accurately detect semantic clones, and it significantly outperforms TBCCD, a state-of-the-art code clone detection tool.
Our results show that \tool significantly outperforms TBCCD showing its effectiveness and generalizing capabilities on unseen datasets.
In the future, we would like to explore the combination of PDG with other program structures like token sequences for learning program representation. We would also like to explore the feasibility of the proposed approach in cross-language clone detection.

\ifCLASSOPTIONcompsoc
  \section*{Acknowledgments}
\else
  \section*{Acknowledgment}
\fi
This work is supported in part by the Department of Science and Technology (DST) (India), Science and Engineering Research Board (SERB), the Confederation of Indian Industry (CII), Infosys Center for Artificial Intelligence at IIIT-Delhi, and Nucleus Software Exports Ltd.

\ifCLASSOPTIONcaptionsoff
  \newpage
\fi

\bibliographystyle{IEEEtran}
\bibliography{list}

\begin{IEEEbiography}
[{\includegraphics[width=1in,height=1.25in,clip,keepaspectratio]{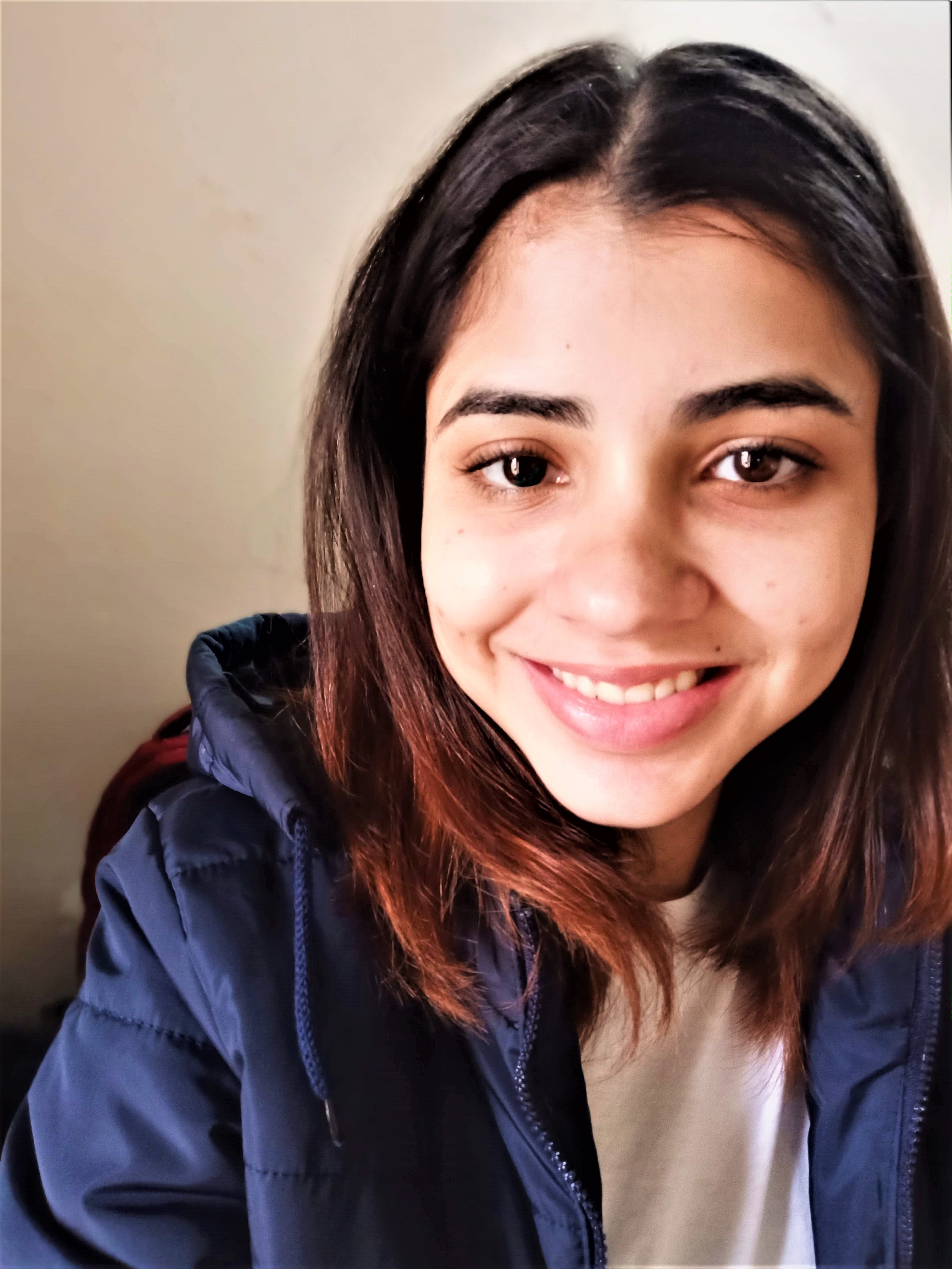}}]{Nikita Mehrotra} is a Ph.D. student in Computer Science and Engineering at
the Indraprastha Institute of Information Technology Delhi (IIIT-Delhi), India. Her research is supported by the Prime Minister’s Research Fellowship with industrial support from Nucleus Software Exports. Her research interests include Deep learning applied software engineering, program understanding, program analysis, software evolution and maintenance.
\end{IEEEbiography}

\begin{IEEEbiography}
[{\includegraphics[width=1in,height=1.25in,clip,keepaspectratio]{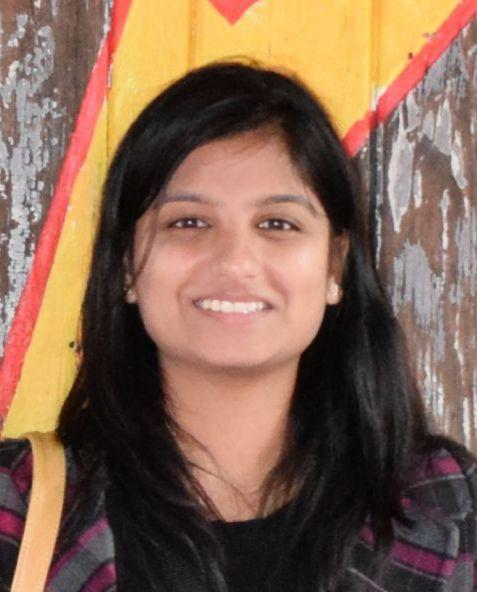}}]
{Navdha Agarwal} is an undergrad student pursuing Computer Science and Applied Mathematics at the Indraprastha Institute of Information Technology Delhi (IIIT-Delhi), India. Her research interests include program analysis and software maintenance.
\end{IEEEbiography}

\begin{IEEEbiography}
[{\includegraphics[width=1in,height=1.25in,clip,keepaspectratio]{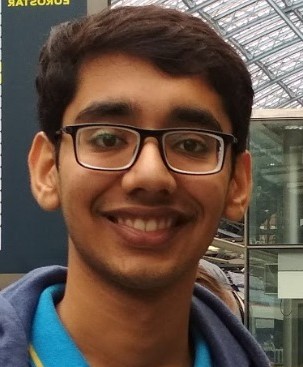}}]
{Piyush Gupta} is an undergrad student pursuing Computer Science Engineering at the Indraprastha Institute of Information Technology Delhi (IIIT-Delhi), India. His research interests include program analysis and software verification.
\end{IEEEbiography}

\begin{IEEEbiography}
[{\includegraphics[width=1in,height=1.25in,clip,keepaspectratio]{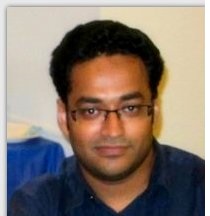}}]{Saket Anand}
is an Assistant Professor at the Indraprastha Institute of Information Technology Delhi (IIIT-Delhi). He received his Ph.D. in Electrical and Computer Engineering from Rutgers University, New Jersey, USA. His research interests include computer vision, machine learning and deep learning. He is a member of the IEEE. 
\end{IEEEbiography}

\begin{IEEEbiography}[{\includegraphics[width=1in,height=1.25in,clip,keepaspectratio]{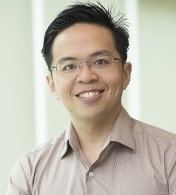}}]{David Lo}
is an Associate Professor in the School of Information Systems, Singapore Management University (SMU). He received his Ph.D. in Computer Science from the National University of Singapore. His research interests include software analytics, software maintenance, empirical software engineering, and cybersecurity.
\end{IEEEbiography}

\begin{IEEEbiography}[{\includegraphics[width=1in,height=1.25in,clip,keepaspectratio]{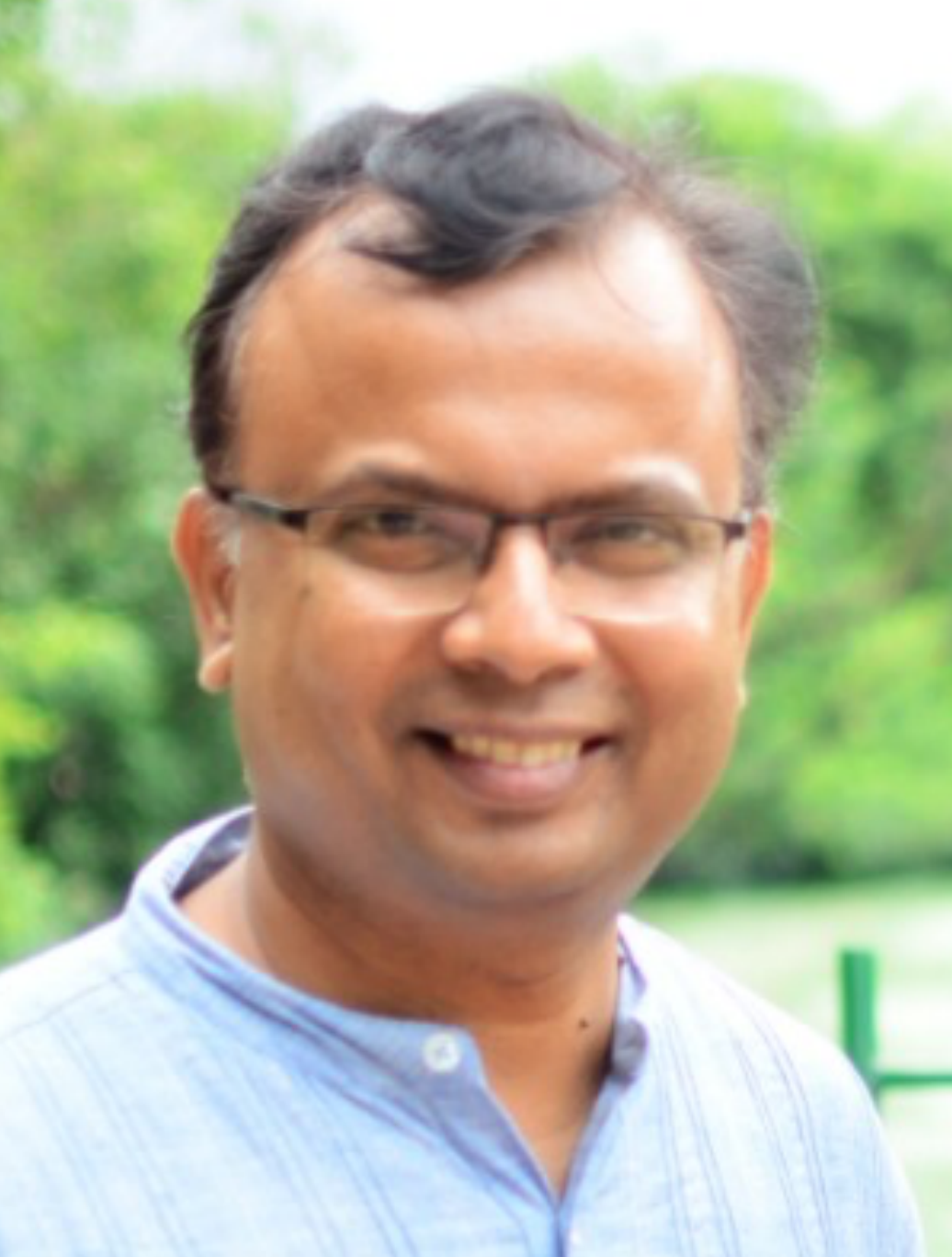}}]{Rahul Purandare}
is an Associate Professor in the department of Computer Science and Engineering at the Indraprastha Institute of Information Technology Delhi (IIIT-Delhi). He received his Ph.D. in Computer Science from the University of Nebraska - Lincoln. His research interests include program analysis, software testing, automatic program repair, code search, and code comprehension.
\end{IEEEbiography}
\end{document}